\documentclass[11pt,a4paper]{article}

\usepackage{jheppub}
\usepackage{epsfig,
latexsym,
amsfonts,
amsmath,
amsthm,
amssymb,
amsbsy,
multirow,
slashed,
color,
mathrsfs,
wasysym,
textcomp,
wrapfig,
bm,
comment}
\usepackage[dvipsnames]{xcolor}
\usepackage{hyperref}
\hypersetup{colorlinks,
	linkcolor = blue,
	anchorcolor = red,
	citecolor = blue,
	filecolor = red,
	urlcolor = blue,
	linktocpage}
\usepackage{caption}
\usepackage{subcaption}

\title{Replica Symmetry Breaking in Bipartite Spin Glasses and Neural Networks}
\author{Gavin S. Hartnett${}^{\,a,b}$,}
\author{Edward Parker${}^{\,c}$,}
\author{Edward Geist${}^{\,a}$}
\affiliation{${}^{a}$The RAND Corporation \\
Santa Monica, CA 90401
}
\affiliation{${}^{b}$STAG Research Centre \\
School of Mathematical Sciences \\
University of Southampton
}
\affiliation{${}^{c}$Department of Physics \\ University of California at Santa Barbara \\
Santa Barbara, CA 93106}
\emailAdd{hartnett@rand.org, tparker@alumni.physics.ucsb.edu, egeist@rand.org}
\abstract{
Some interesting recent advances in the theoretical understanding of neural networks have been informed by results from the physics of disordered many-body systems. Motivated by these findings, this work uses the replica technique to study the mathematically tractable bipartite Sherrington-Kirkpatrick (SK) spin glass model, which is formally similar to a Restricted Boltzmann Machine (RBM) neural network. The bipartite SK model has been previously studied assuming replica symmetry; here this assumption is relaxed and a replica symmetry breaking analysis is performed. The bipartite SK model is found to have many features in common with Parisi's solution of the original, unipartite SK model, including the existence of a multitude of pure states which are related in a hierarchical, ultrametric fashion. As an application of this analysis, the optimal cost for a graph partitioning problem is shown to be simply related to the ground state energy of the bipartite SK model. As a second application, empirical investigations reveal that the Gibbs sampled outputs of an RBM trained on the MNIST data set are more ultrametrically distributed than the input data itself.
}

\begin{document}
\maketitle

\section{Introduction}
Research into deep neural networks has advanced at a stunning rate in recent years, leading to tremendous progress for a number of difficult machine learning problems and benchmarks \cite{lecun2015deep}. Despite these impressive accomplishments, there is a sense in which much of the recent progress has been made in engineering new methods and techniques, and that the development of a robust theoretical understanding has lagged behind. This state of affairs is a natural consequence both of the inherent difficulty in building a strong theoretical framework, and in the unprecedented rate of success that neural networks have had in solving machine learning problems. 

Recently, some interesting advances in improving the theoretical understanding of neural networks have been inspired by results from physics, particularly the physics of disordered many-body systems. For example, in \cite{bray2007statistics}, Bray and Dean considered Gaussian random fields in high-dimensional spaces and calculated many properties of the distribution of critical points. Although the neural networks used in practical applications are certainly not Gaussian random fields, Dauphin \textit{et al.}\ \cite{dauphin2014identifying} found through an empirical analysis that Bray and Dean's observations for Gaussian random fields hold for neural networks trained on common machine learning data sets, such as MNIST and CIFAR-10. These include the result that high-error saddles pose a difficulty for training neural networks, and that most local minima have errors very close to the global minimum error. By drawing inspiration from physics, machine learning researchers were able to improve their understanding of neural networks, and moreover were led to propose a new Newton's method algorithm for optimization problems which suffer from many saddle points. 

Another advance in the theoretical understanding of neural networks that borrows from physics was made in \cite{choromanska2015loss}, where it was observed that under certain simplifying assumptions, the error surface of feed-forward neural networks can be mapped precisely to the energy landscape for a physics model known as the $p$-spin spherical spin-glass, for which exact results have recently been derived in \cite{auffinger2013random, auffinger2013complexity}. As in the Dauphin \textit{et al.}\ study, these authors then empirically verified that these mathematical insights qualitatively apply to neural networks trained on MNIST, even though the assumptions behind the spin glass analysis no longer hold.\footnote{It is also worth noting that interesting progress has been made recently in the other direction, namely using machine learning methods to learn more about condensed matter systems \cite{carrasquilla2017machine}.} Yet another recent work along these lines is \cite{schoenholz2017correspondence}.

Inspired by these results, in this work we study in detail a mathematically tractable spin glass model, and then use these results to better understand a family of neural networks. The spin glass model we consider is a simple extension of the paradigmatic Sherrington-Kirkpatrick (SK) model \cite{sherrington1975solvable}. The SK model is an infinite-range spin glass with Hamiltonian
\begin{equation}
\label{eq:SKHamiltonian}
H_{\text{SK}} = - \sum_{i<j} J_{ij} s_i s_j \,.
\end{equation}
Here $s_i$, $i=1,...,N$ are binary spins, $s_i \in \{-1,1\}$. The 2-spin couplings $J_{ij}$ are independent identically distributed (iid) random variables drawn from a normal distribution with mean $J_0$ and standard deviation $J$. The SK model is an ideal toy model; despite its simplicity its theoretical analysis is very rich and involves phenomena such as a continuous order parameter, an ultrametric space of states, and stochastic branching processes. In fact, the historical analysis of the SK model is very closely tied to the development of spin glass theory.

As every spin in the SK model couples to every other spin, the couplings may be characterized in terms of a complete graph. In this paper, we consider the extension of the SK model to complete bipartite graphs, which we will refer to as the ``bipartite SK model.'' The special case of \emph{balanced} bipartite graphs has been studied for quite some time \cite{korenblit1985spin, korenblit1985magnetic, korenblit1987magnetic, fyodorov1987antiferromagnetic, fyodorov1987phase, oppermann2004modulated}, but the general case has only been studied much more recently \cite{barra2018phase, barra2017phase2, barra2011equilibrium, barra2012glassy, barra2014mean, barra2015multi, panchenko2015free, genovese2016non, fedele2012rigorous, genovese2012universality}. These papers have calculated the model's phase diagram in certain regimes \cite{fyodorov1987antiferromagnetic, barra2011equilibrium, barra2018phase} (related to our results in the next section), but to our knowledge the replica symmetry breaking of the bipartite SK model, which is crucial for correctly describing the physics of the spin glass phase, has been much less well-studied.\footnote{Some relevant works are as follows. \cite{oppermann2004modulated} considered the replica symmetry breaking for the special case of balanced bipartite graphs. \cite{barra2015multi} considered the replica symmetry breaking of multi-species spin models and proved an upper bound on the free energy, and \cite{panchenko2015free} subsequently showed that the bound is sharp. However, the model they considered necessarily allows self-interactions between species, unlike the bipartite model considered here. \cite{baik2017free} considered the bipartite spherical SK model. \cite{decelle2017spectral, decelle2018thermodynamics} performed a replica-symmetric analysis very similar to the one we perform in Sec.~\ref{sec:rs_analysis}.}

Our primary motivation for studying the bipartite SK model is its similarity to a family of neural networks known as Restricted Boltzmann Machines (RBMs) \cite{smolensky1986information}. RBMs are generative models used in unsupervised learning, and are appealing to study from a physics perspective because they are energy-based models with a Hamiltonian of the same parametric form as the bipartite SK model. In fact, the only real difference between the bipartite SK model and the RBM lies in how the parameters are chosen. In the bipartite SK model they are iid normally distributed, whereas in the RBM they are determined by a learning algorithm. Therefore, similar in spirit to Dauphin et al \cite{dauphin2014identifying} and Choromanska et al \cite{choromanska2015loss}, it is plausible that the spin glass analysis of the bipartite SK model could provide novel insights into RBMs. Indeed, we will find this to be the case.

The outline of this paper is as follows. First, in Sec.~\ref{sec:bipartiteSK} we analyze the bipartite SK model, considering both the replica symmetric and replica symmetry breaking cases. Then in Sec.~\ref{sec:optimization} we consider a combinatorial optimization problem for which the optimal cost can be shown to be simply related to the ground state energy of the bipartite SK model. In Sec.~\ref{sec:mcmc} we present the results from a range of numerical investigations, including Markov-chain Monte Carlo (MC) simulations, and find good agreement. In Sec.~\ref{sec:RBM}, we empirically investigate the extent to which RBMs trained on realistic data exhibit spin glass phenomena. We conclude in Sec.~\ref{sec:conclusion}.

\section{The Bipartite Sherrington-Kirkpatrick Model \label{sec:bipartiteSK}}
In this section we study the statistical physics of an infinite-range spin glass model called the bipartite SK model, which generalizes the famous Sherrington-Kirkpatrick (SK) model \cite{sherrington1975solvable} to a bipartite coupling graph. The Hamiltonian is
\begin{equation}
\label{eq:bipartiteSK}
H = - \sum_{i=1}^{N_v} \sum_{j=1}^{N_h} W_{ij} v_i h_j - \sum_{i=1}^{N_v} b_i^{(v)} v_i - \sum_{j=1}^{N_h} b_j^{(h)} h_j \,.
\end{equation}
Here $v_i, h_j \in \{-1, 1\}$ are spin variables which, anticipating the connection with RBMs in Sec.~\ref{sec:RBM}, we will call the visible and hidden spins, respectively. There are $N$ spins in total, divided into $N_v$ visible and $N_h$ hidden spins. We will use the convention that the index $i$ runs from 1 to $N_v$, and the index $j$ runs from $1$ to $N_h$. The parameters of the model are an $N_v \times N_h$ matrix of couplings $W_{ij}$, and the $N_v$- and $N_h$-length vectors $b_i^{(v)}$ and $b_j^{(h)}$ which correspond to an independent external magnetic field for each spin. Following the machine learning literature, we will also refer to these as the biases. The Hamiltonian defines a Gibbs probability distribution over the spin variables, \begin{equation}
\label{eq:Gibbs}
p(\bm{v}, \bm{h}) = \frac{e^{-\beta H}}{Z} \,,
\end{equation}
with $\beta = 1/T$ the inverse temperature and 
\begin{equation}
Z = \text{Tr } e^{-\beta H} 
\end{equation} 
the partition function. The $\text{Tr}$ symbol indicates a sum over all $2^{N_v + N_h} = 2^N$ configurations of the spin variables. The sum is intractable for both sets of spins together, but because there are no visible-visible or hidden-hidden couplings, it can be performed analytically if one set of spins is held fixed. For example, the marginal distribution over the visible spins can be written as $p(\bm{v}) = e^{-\beta H_v}/Z$, with
\begin{equation}
\label{eq:RBMvisibleHamiltonianSpin}
H_v = - \sum_i b_i^{(v)} v_i - \frac{1}{\beta} \sum_j \ln \left[ 2 \cosh \left( \beta \, b_j^{(h)} + \beta \sum_i v_i W_{ij} \right) \right] \,,
\end{equation}
and similarly for the hidden spins.

In the bipartite SK model, the parameters appearing in the Hamiltonian are independently drawn from normal distributions,
\begin{equation}
\label{eq:gaussian_parameters} 
W_{ij} \sim \mathcal{N}(W_0, W) \,, \qquad b_{i}^{(v)} \sim \mathcal{N}\left(b_0^{(v)}, b^{(v)}\right) \,, \qquad b_{j}^{(h)} \sim \mathcal{N}\left(b_0^{(h)}, b^{(h)}\right)
\end{equation}
where the notation $x \sim \mathcal{N}(\mu, \sigma)$ indicates that the random variable $x$ is drawn from a normal distribution with mean $\mu$ and standard deviation $\sigma$. Our motivation for studying the bipartite model is its similarity to a class of neural networks used for unsupervised learning known as Restricted Boltzmann Machines (RBMs). RBMs are probabilistic models whose Hamiltonians take the same general form as Eq.~\ref{eq:bipartiteSK}, but with the parameters allowed to take on arbitrary values.\footnote{\label{ftnt:KS}In the symmetric, uniform-bias case $N_v = N_h$, $b_0^{(v)} = b_0^{(h)}$, $b^{(v)} = b^{(h)} = 0$, the bipartite SK model is known as the Korenblit-Shender model \cite{korenblit1985spin, korenblit1985magnetic}. $N_v$ and $N_h$ are typically equal in physical spin systems but not in RBMs.} The spin-glass properties of RBMs will be the focus of Sec.~\ref{sec:RBM}. 

Calculating the free energy
\begin{equation}
F = -\beta^{-1} \ln Z
\end{equation}
for a specific choice of parameters $W_{ij}, b_i^{(v)}, b_i^{(h)}$ (known as a ``disorder realization" in physics terminology) is an intractable problem, but in the large-$N$ limit we can calculate the expectation value of the free energy with respect to the probability distributions given in Eq.'s~\ref{eq:gaussian_parameters}. We will use the notation $[X]$ to denote this expectation value (known as the ``disorder-averaged value'') for any quantity $X$.

Crucially, in the large-$N$ limit, certain ``self-averaging" quantities (including the free energy) become independent of the particular disorder realization. In order to have a well-defined large-$N$ limit in which the number of spins is infinite, the moments of the 2-spin coupling $W_{ij}$ will need to be rescaled. Defining ${W_0 = w_0/\sqrt{N_v N_h}}$ and ${W = w/\sqrt[4]{N_v N_h}}$, the $N_v, N_h \rightarrow \infty$ limit may be taken while keeping $w_0$, $w$ finite. We will also find it useful to introduce
\begin{equation}
\alpha_v = \frac{N_v}{N}, \qquad \alpha_h = \frac{N_h}{N}.
\end{equation}

We can calculate $[F]$ using the replica method, which is based on the simple mathematical identity 
\begin{equation}
\label{eq:replicatrick}
[\ln Z] = \frac{\partial [Z^n]}{\partial n} \Bigg|_{n = 0} \,.
\end{equation}
$[Z^n]$ is initially computed with $n$ taken to be a positive integer, and each copy of $Z$ is referred to as a replica. At the end of the calculation, the above formula is used to compute $[\ln Z]$ and hence also $[F]$. There are many mathematical subtleties associated with the replica method which have been discussed extensively in the physics literature; we therefore omit any further discussion here.

For the bipartite SK model, the disorder average may be taken immediately, yielding
\begin{align}
\label{eq:Zbeforeintegralxform}
[ Z^n ] = \text{Tr}_n\, \exp& \left[\bar{W}_0 \sum \limits_{a, i, j} v_i^{a} h_j^{a} + \frac{\bar{W}^2}{2} \sum \limits_{a, b, i, j} v_i^{a} v_i^{b} h_j^{a} h_j^{b} \right. \\
&\left. + \bar{b}_0^{(v)} \sum \limits_{a, i} v_i^{a} + \frac{\left(\bar{b}^{(v)}\right)^2}{2} \sum_{a, b, i} v_i^{a} v_i^{b} + \bar{b}_0^{(h)} \sum \limits_{a, j} h_j^{a} + \frac{\left(\bar{b}^{(h)}\right)^2}{2} \sum_{a, b, j} h_j^{a} h_j^{b}\right] \nonumber
\end{align}
Here the indices $a, b$, run from 1 to $n$, and label each replica. $\text{Tr}_n$ indicates a trace over all replicas, and a bar indicates that the inverse temperature $\beta$ has been absorbed into the parameter, for example $\bar{W}_{ij} = \beta\, W_{ij}$. 

In the analysis of the original, unipartite SK model, the Hubbard-Stratonovich integral transform is used to convert the argument of the exponential to be quadratic in the spin variables, at the cost of introducing an integral. For the bipartite SK model, there are two species of spin variables, and a different integral transform is required. Such a transform is provided by
\begin{equation}
\label{eq:integralxform}
e^{\frac{B C}{\sqrt{2} a}} = \frac{a^2}{2^{3/2}\pi^2} \int_{-\infty}^{\infty} \mathrm{d} x\, \mathrm{d} \tilde{x}\,  \mathrm{d} y\,  \mathrm{d} \tilde{y}\,  e^{-a\left(x^2 - \sqrt{2} x y + y^2 + \frac{1}{2}\tilde{x}^2 + \frac{1}{2}\tilde{y}^2\right) + B (x + i \tilde{x}) + C (y + i \tilde{y})} \,,
\end{equation}
which converges for $a>0$. This transform may be applied to the $\bar{W}_0$ term in Eq.~\ref{eq:Zbeforeintegralxform} for
\begin{align}
& a = N \sqrt{\frac{\alpha_v \alpha_h}{2}} \bar{w}_0 \,, \quad B =  \bar{w}_0 \sum_i v_i^{a} \,, \quad C = \bar{w}_0 \sum_j h_j^{a} \,,
\end{align}
and to the $\bar{W}^2$ term for
\begin{align}
& a =  N \sqrt{\frac{\alpha_v \alpha_h}{2}} \frac{\bar{w}^2}{2} \,, \quad B = \frac{\bar{w}^2}{2} \sum_i v_i^{a} v_i^{b} \,, \quad C = \frac{\bar{w}^2}{2} \sum_j h_j^{a} h_j^{b} \,.
\end{align}
The resulting expression is 
\begin{equation}
\label{eq:Zafterintegralxform}
[Z^n] = \left( \int \prod_{a < b} \mathrm{d} X_{a b} \mathrm{d}  \tilde{X}_{a b} \mathrm{d}  Y_{a b} \mathrm{d}  \tilde{Y}_{a b} \right) \left( \int \prod_{a} \mathrm{d} U_{a} \mathrm{d}  \tilde{U}_{a} \mathrm{d}  V_{a} \mathrm{d}  \tilde{V}_{a} \right) e^{-\beta N n \mathcal{F}_n} \,,
\end{equation}
where
\begin{align}
\beta n \mathcal{F}_n =&\ \bar{w}^2 \sqrt{\alpha_v \alpha_h} \sum_{a<b} \left( \frac{X_{a b}^2}{\sqrt{2}} - X_{a b} Y_{a b} + \frac{Y_{a b}^2}{\sqrt{2}} + \frac{\tilde{X}_{a b}^2}{2\sqrt{2}}  +  \frac{\tilde{Y}_{a b}^2}{2\sqrt{2}} \right) \\
& + \bar{w_0}  \sqrt{\alpha_v \alpha_h} \sum_{a} \left( \frac{U_{a}^2}{\sqrt{2}} - U_{a} V_{a} + \frac{V_{a}^2}{\sqrt{2}} + \frac{\tilde{U}_{a}^2}{2\sqrt{2}} + \frac{\tilde{V}_{a}^2}{2\sqrt{2}} \right) \nonumber \\
& - \frac{n}{2} \left(\sqrt{\alpha_v \alpha_h} \bar{w}^2 +  \alpha_v \left(\bar{b}^{(v)}\right)^2 + \alpha_h \left(\bar{b}^{(h)}\right)^2 \right) - \alpha_v \log \text{Tr}_{n,v} \, \Psi_v - \alpha_h \log \text{Tr}_{n,h} \, \Psi_h\,, \nonumber
\end{align}
and
\begin{subequations}
\begin{equation}
\Psi_v = \exp \left[\sum_{a<b} \left( \bar{w}^2 (X_{ab} + i \tilde{X}_{ab}) + \left(\bar{b}^{(v)}\right)^2 \right) v^{a} v^{b}  + \sum_{a} \left( \bar{w}_0 (U_{a} + i \tilde{U}_{a}) + \bar{b}_0^{(v)} \right) v^{a}\right] \,,
\end{equation}
\begin{equation}
\Psi_h = \exp \left[\sum_{a<b} \left( \bar{w}^2 (Y_{ab} + i \tilde{Y}_{ab}) + \left(\bar{b}^{(h)}\right)^2 \right) h^{a} h^{b}  + \sum_{a} \left(\bar{w}_0 (V_{a} + i \tilde{V}_{a}) + \bar{b}_0^{(h)} \right) h^{a}\right] \,.
\end{equation}
\end{subequations}
In the above, the site indices $i,j$ for the spins have been dropped because the sums over sites factorize and each site is treated equally due to the disorder average, and $\text{Tr}_{n,v}$ refers to a trace over all $n$ replicas of just the visible spins only, and similarly for $\text{Tr}_{n,h}$.

In the large-$N$ limit, the integral may be evaluated by steepest descents: ${ [ Z^n ] \approx e^{-\beta N n \mathcal{F}_n{}^*} }$. The $*$ indicates that $\mathcal{F}_n$ has been evaluated at a saddle, which is determined by the following equations:
\begin{subequations}
\label{eq:saddleeqns}
\begin{equation}
X^*_{a b} =  \sqrt{2} \sqrt{\frac{\alpha_v}{\alpha_h}} q_{a b}^{(v)} + \sqrt{\frac{\alpha_h}{\alpha_v}} q_{a b}^{(h)} \,, \qquad \tilde{X}^*_{a b} = i \sqrt{2} \sqrt{\frac{\alpha_v}{\alpha_h}} q_{a b}^{(v)} \,,
\end{equation}
\begin{equation}
Y_{a b}^* = \sqrt{\frac{\alpha_v}{\alpha_h}} q_{a b}^{(v)} + \sqrt{2}\sqrt{\frac{\alpha_h}{\alpha_v}} q_{a b}^{(h)} \,, \qquad \tilde{Y}_{a b}^* = i \sqrt{2} \sqrt{\frac{\alpha_h}{\alpha_v}} q_{a b}^{(h)} \,,
\end{equation}
\begin{equation}
U^*_{a} = \sqrt{2} \sqrt{\frac{\alpha_v}{\alpha_h}}m_{a}^{(v)} + \sqrt{\frac{\alpha_h}{\alpha_v}} m_{a}^{(h)} \,, \qquad \tilde{U}^*_{a} = i \sqrt{2} \sqrt{\frac{\alpha_v}{\alpha_h}} m_{a}^{(v)} \,,
\end{equation}
\begin{equation}
V_{a}^* =\sqrt{\frac{\alpha_v}{\alpha_h}} m_{a}^{(v)} + \sqrt{2} \sqrt{\frac{\alpha_v}{\alpha_h}} m_{a}^{(h)} \,, \qquad \tilde{V}_{a} = i \sqrt{2} \sqrt{\frac{\alpha_h}{\alpha_v}} m_{a}^{(h)} \,,
\end{equation}
\end{subequations}
where we have introduced the average magnetizations
\begin{equation}
m_{a}^{(v)} = \frac{1}{N_v} \sum_i  [ \langle  v_i^{a} \rangle ] \,, \qquad m_{a}^{(h)} = \frac{1}{N_h} \sum_j [ \langle h_j^{a} \rangle ] \,,
\end{equation}
as well as the overlaps
\begin{equation}
q_{a b}^{(v)} = \frac{1}{N_v} \sum_i [ \langle  v_i^{a}v_i^{b} \rangle ] \,, \qquad q_{ab}^{(h)} = \frac{1}{N_h} \sum_j [ \langle h_j^{a}h_j^{b} \rangle ] \,, \qquad \text{for } a \neq b \,.
\end{equation}
Angled brackets $\langle \cdot \rangle$ denote a thermal average with respect to the Gibbs distribution Eq.~\ref{eq:Gibbs}.

It is not useful to immediately impose the saddle point equations because the trace must still be carried out in the $\Psi_{v,h}$ terms. Before carrying out the traces, however, it is convenient to first convert to complex variables, defined by 
\begin{subequations}
\label{eq:complexvariables}
\begin{equation}
M_{a}^{(h)} = \sqrt{\frac{\alpha_v}{\alpha_h}} \left( U_{a} + i \, \tilde{U}_{a} \right) \,, \qquad M_{a}^{(v)} = \sqrt{\frac{\alpha_h}{\alpha_v}} \left( V_{a} + i \, \tilde{V}_{a } \right) \,,
\end{equation}
\begin{equation}
\widehat{M}_{a}^{(h)} = \sqrt{\frac{\alpha_v}{\alpha_h}} \left( U_{a} - i \, \tilde{U}_{a} \right) \,, \qquad \widehat{M}_{a}^{(v)} = \sqrt{\frac{\alpha_h}{\alpha_v}} \left( V_{a} - i \, \tilde{V}_{a } \right) \,,
\end{equation}
\begin{equation}
Q_{a b}^{(h)} = \sqrt{\frac{\alpha_v}{\alpha_h}} \left( X_{a b} + i \, \tilde{X}_{a b} \right) \,, \qquad Q_{a b}^{(v)} = \sqrt{\frac{\alpha_h}{\alpha_v}} \left( Y_{a b} + i \, \tilde{Y}_{a b} \right) \,,
\end{equation}
\begin{equation}
\widehat{Q}_{a b}^{(h)} = \sqrt{\frac{\alpha_v}{\alpha_h}} \left( X_{a b} - i \, \tilde{X}_{a b} \right) \,, \qquad \widehat{Q}_{a b}^{(v)} = \sqrt{\frac{\alpha_h}{\alpha_v}} \left( Y_{a b} - i \, \tilde{Y}_{a b} \right) \,,
\end{equation}
\end{subequations}
These variables have the useful property that the hatted variables do not depend on the traces, and so their saddle equations may be immediately solved for, yielding\footnote{Note that the hatted quantities are not necessarily the complex conjugates of the un-hatted quantities: this is true only if the saddle solutions of the variable $X_{a b},\tilde{X}_{a b}, Y_{a b}, \tilde{Y}_{a b}$ are all real, which will turn out not to be the case.}
\begin{align}
\label{eq:free_energy_no_simplification}
\beta n \mathcal{F}_n^* =&\ \sqrt{\alpha_v \alpha_h} \left(\bar{w}^2 \sum_{a < b}  Q_{ab}^{(v)} \,Q_{ab}^{(h)} + \bar{w}_0 \sum_{a} M_{a}^{(v)} \,M_{a}^{(h)} \right) \\
& - \frac{n}{2} \left(\sqrt{\alpha_v \alpha_h} \bar{w}^2 +  \alpha_v \left(\bar{b}^{(v)}\right)^2 + \alpha_h \left(\bar{b}^{(h)}\right)^2 \right) - \alpha_v \log \text{Tr}_{n,v} \, \Psi_v - \alpha_h \log \text{Tr}_{n,h} \, \Psi_h \nonumber \,,
\end{align}
where now
\begin{equation}
\Psi_v = \exp \left[\sum \limits_{a<b} \left( \sqrt{\frac{\alpha_h}{\alpha_v}} \bar{w}^2 Q_{ab}^{(h)} + \left(\bar{b}^{(v)}\right)^2 \right) v^{a} v^{b}  + \sum_{a} \left( \sqrt{\frac{\alpha_h}{\alpha_v}} \bar{w}_0 M_a^{(h)} + \bar{b}_0^{(v)} \right) v^{a}\right] \,, \quad v \leftrightarrow h.
\end{equation}
(The notation $v \leftrightarrow h$ means that the equivalent equation with every appearance of the labels $v$ and $h$ interchanged also holds.) This choice of variables has the additional property that the saddle equations Eq.~\ref{eq:saddleeqns} become simply
\begin{equation}
\label{eq:saddlepointeqs}
M_{a}^{(v)}{}^* = m_{a}^{(v)} \,, \qquad Q_{a b}^{(v)}{}^* = q_{a b}^{(v)} \,, \qquad v \leftrightarrow h.
\end{equation}

At this point the calculation has been formulated as a simple extension of the usual replica symmetry breaking calculation of the unipartite model -- see for example \cite{dotsenko2005introduction, mezard1987spin, nishimori2001statistical}. One interesting point to note is that if $\alpha_v = \alpha_h = 1/2$ and $b_0^{(v)} = b_0^{(h)}$, $b^{(v)} = b^{(h)}$, then there is a symmetry between the visible and hidden spins and the expression reduces to exactly that of the unipartite SK model. We have verified that our results reduce to those of the SK model when this condition is imposed. 

For simplicity, we will henceforth neglect the bias terms by setting them to zero.\footnote{\cite{korenblit1985spin, korenblit1985magnetic, korenblit1987magnetic, fyodorov1987antiferromagnetic, fyodorov1987phase, oppermann2004modulated} considered the effect of a uniform bias in the special case of the Korenblit-Shender model defined in footnote \ref{ftnt:KS}.} The free energy is invariant under the transformation $W_{ij} \to -W_{ij},\ b_i^{(v)} \to -b_i^{(v)}$, which corresponds to flipping all the spins in the visible part (and also under the equivalent transformation for the hidden part). In the bipartite SK model, this corresponds to ${w_0 \to -w_0}$ and either ${b_0^{(v)} \to -b_0^{(v)}}$ or ${b_0^{(h)} \to -b_0^{(h)}}$ but not both). In the zero-bias case we will consider, the free energy is therefore symmetric in $w_0$, and without loss of generality we will only consider the ferromagnetic case $w_0 > 0$.

As in the SK model, an ansatz for the matrix structure of the integration variables will need to be specified in order to proceed. We will first assume a replica symmetric ansatz and analyze the phase diagram. Replica symmetry is broken in the spin glass phase, and in the subsequent section the replica symmetry breaking (RSB) ansatz of Parisi will be used. 

\subsection{Replica Symmetric Analysis \label{sec:rs_analysis}}
First, we assume a replica symmetric ansatz, with all matrix and vector entries equal: 
\begin{equation}
Q_{a b}^{(h)} {}^* = q^{(h)} \,, \quad Q_{a b}^{(v)} {}^* = q^{(v)} \,, \quad M^{(h)}_{a}{}^* = m^{(h)} \,, \quad M^{(v)}_{a}{}^* = m^{(v)} \,,
\end{equation}
which results in
\begin{align}
\beta \mathcal{F}_n^* = \sqrt{\alpha_v \alpha_h} \left[ \frac{\bar{w}^2}{2} \Big( (n-1) q^{(v)} q^{(h)} - 1 \Big) + \bar{w}_0 \, m^{(v)} m^{(h)} \right]- \frac{\alpha_v}{n} \log \text{Tr}_{n,v} \, \Psi_v - \frac{\alpha_h}{n} \log \text{Tr}_{n,h} \, \Psi_h \,,
\end{align}
where now
\begin{equation}
\Psi_v = \exp \left[ \sqrt{\frac{\alpha_h}{\alpha_v}}\left(\bar{w}^2 q^{(h)} \sum_{a < b} v^{a} v^{b} + \bar{w}_0 \, m^{(h)} \sum_{a} v^{a} \right) \right] \,, \qquad v \leftrightarrow h.
\end{equation}

The quadratic spin terms in $\Psi_{v,h}$ may be linearized through the Hubbard-Stratonovich transform, after which the traces become trivial. Finally, Eq.~\ref{eq:replicatrick} may be used to yield the free energy density, $[f] = [F]/N = \mathcal{F}_0^*$, with the result that
\begin{align}
\label{eq:free_energy_RS}
\beta [ f ] =& -\frac{\sqrt{\alpha_v \alpha_h} \bar{w}^2}{2} \left( q^{(v)} - 1 \right)\left( q^{(h)}- 1 \right) + \sqrt{\alpha_v \alpha_h} \bar{w_0} \, m^{(v)} m^{(h)} \\
& - \alpha_v \langle \log( 2 \cosh H_h(z)) \rangle_z - \alpha_h \langle \log(2 \cosh H_v(z))\rangle_z \nonumber \,,
\end{align}
where 
\begin{equation}
H_v(z) = \bar{w} \left(\frac{\alpha_v}{\alpha_h}\right)^\frac{1}{4} \sqrt{q^{(v)}} z + \bar{w}_0 \sqrt{\frac{\alpha_v}{\alpha_h}} m^{(v)} \,, \quad v \leftrightarrow h \,.
\end{equation}
The notation $\langle \cdot \rangle_z$ indicates an expectation value taken over the Gaussian random variable $z$ with zero mean and unit variance that was introduced by the Hubbard-Stratonovich transform,
\begin{equation}
\langle g(z) \rangle_z = \int_{-\infty}^{\infty} \frac{\mathrm{d} z}{\sqrt{2\pi}} e^{-z^2/2} g(z) \,.
\end{equation}
Extremizing the free energy with respect to the variables $\big\{ m^{(v)}, m^{(h)}, q^{(v)}, q^{(h)}\big\}$ yields the saddle equations
\begin{subequations}
\label{eq:RSsaddle}
\begin{equation}
\label{eq:RSsaddleM}
m^{(v)} = \left\langle \tanh H_h(z) \right\rangle_z \,, \quad v \leftrightarrow h \,,
\end{equation}
\begin{equation}
\label{eq:RSsaddleQ}
q^{(v)} = \left\langle \tanh^2 H_h(z) \right\rangle_z \,, \quad v \leftrightarrow h \,.
\end{equation}
\end{subequations}

We can now map out the phase diagram of the zero-bias system, and we find three phases. In the paramagnetic phase, with $m = q = 0$, each spin fluctuates randomly over time. In the ferromagnetic phase, with $q = m^2 \neq 0$, the spins are frozen into a single globally aligned symmetry-breaking direction. In the spin-glass phase, with $m = 0$ but $q \neq 0$, each individual spin is frozen in a particular direction, but an equal number are frozen in each direction, so there is no net magnetization.

\begin{figure}
\centering
\begin{subfigure}{0.5\textwidth}
  \centering
\includegraphics[width=1.0\textwidth]{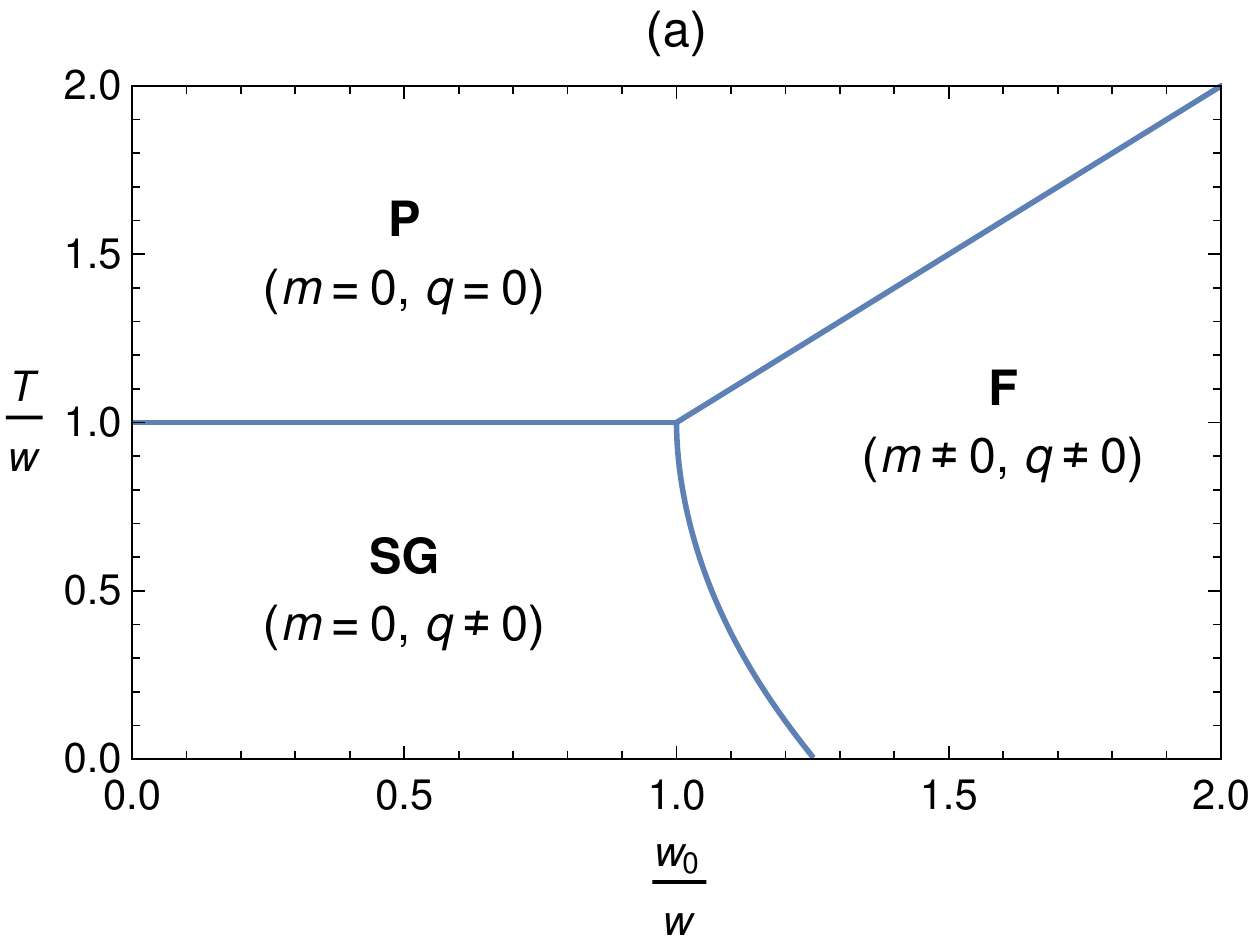}  
\end{subfigure}%
\begin{subfigure}{0.5\textwidth}
  \centering
\includegraphics[width=1.0\textwidth]{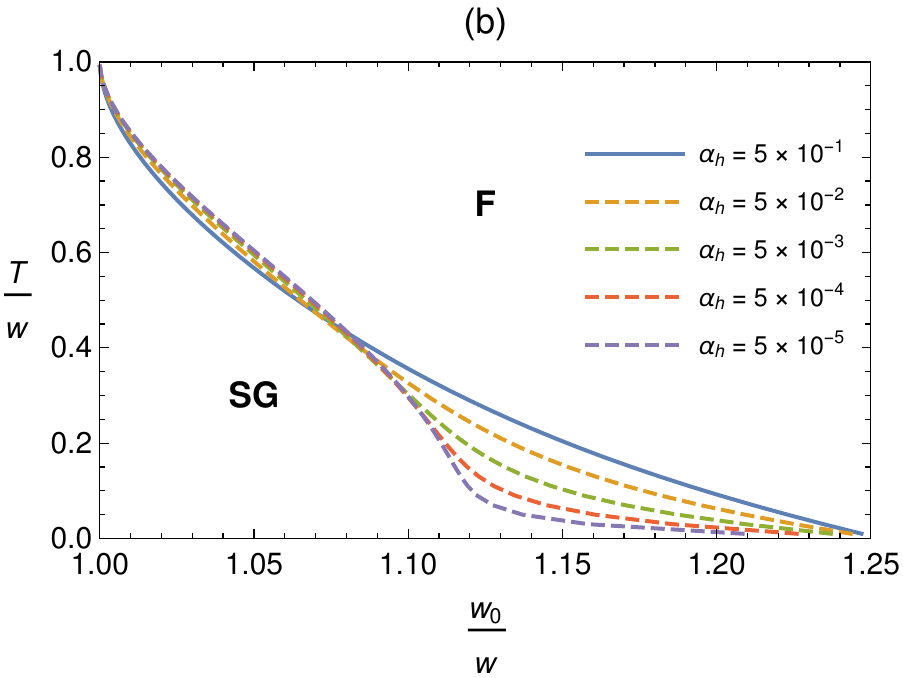}
\end{subfigure}
\caption{\label{fig:phasediagram} (a): The phase diagram for the bipartite SK model (restricted to zero biases), as predicted by the replica-symmetric analysis. Only the SG/F phase boundary depends on the relative fractions of visible and hidden spins, and in this plot we have set $\alpha_{v} = \alpha_{h} = 1/2$ so that this phase diagram is exactly the same as in the SK model. (b): To display the effect of a disparity between the numbers of the two types of spins, we have zoomed into the SG/F transition for various values of $\alpha_h$. The solid line corresponds to $\alpha_v = \alpha_h = 1/2$, and starting in the lower right-hand corner, the curves descend as $\alpha_h$ decreases.
}
\end{figure}

The equations Eq.~\ref{eq:RSsaddle} must be solved numerically, except in the vicinity of the transitions that border the paramagnetic phase, at which both the magnetizations and overlaps are small and a perturbative analysis may be applied. The ferromagnetic/spin glass transition is \textit{not} accessible via perturbative methods as the overlap is non-zero on both sides of the transition. Perturbative analysis yields the following critical lines: $\bar{w}_0 = 1$ for $0 \le \bar{w} < 1$ as a phase boundary between the paramagnetic and ferromagnetic phases, and $\bar{w} = 1$ as the boundary between the paramagnetic and spin glass phases. The ferromagnet/spin glass phase boundary can be solved numerically by expanding the equations Eq.~\ref{eq:RSsaddle} around zero magnetization. The paramagnetic/spin glass phase boundary for $\bar{w}_0 =0$ was previously derived (with different conventions) in \cite{barra2011equilibrium, barra2018phase}. We plot the zero-bias phase diagram in Fig.~\ref{fig:phasediagram}.\footnote{\label{ftnt:biases}This phase diagram only applies for the case of zero biases. When the bias means $b_0^{(v,h)}$ are non-zero, the Hamiltonian is no longer symmetric under the transformation $v_i \rightarrow - v_i,\ h_j \rightarrow -h_j$ and the ferromagnetic and antiferromagnetic cases are no longer equivalent. Assuming the bias means have the same sign, in the ferromagnetic case $w_0 > 0$ the transition between the paramagnetic and ferromagnetic phases disappears and they merge into a single phase, which has $m \neq 0$ but is best thought of as the paramagnetic phase, because the ground state is unique and there is no symmetry breaking. In the antiferromagnetic case $w_0 < 0$, the two parts magnetize unequally, with the larger part more strongly magnetized (in the symmetric case $\alpha_v = \alpha_h$, there remains a $\mathbb{Z}_2$ symmetry-breaking phase in which the larger magnetization is selected spontaneously). Both cases still have a spin glass phase characterized by many pure states \cite{mezard1987spin, korenblit1985spin}, but with a net magnetization.}

In more detail, the ferromagnet/spin glass boundary $\bar{w}_0(\bar{w})$ may be solved for by first specifying a value of $\bar{w}$. Then the overlaps are determined by solving Eq.'s~\ref{eq:RSsaddleQ} evaluated at zero magnetization. Finally, the value of $\bar{w}_0$ corresponding to the phase boundary may be determined by solving Eq.'s~\ref{eq:RSsaddleM} expanded to linear order in the magnetizations. The final equation governing the phase boundary is
\begin{equation}
1 - \bar{\omega_0}^2 \left\langle \text{sech}^2 \left[\left(\frac{\alpha_h}{\alpha_v}\right)^\frac{1}{4} \sqrt{q^{(h)}} \bar{w} z \right] \right\rangle_z \left\langle \text{sech}^2 \left[\left(\frac{\alpha_v}{\alpha_h}\right)^\frac{1}{4} \sqrt{q^{(v)}} \bar{w} z \right] \right\rangle_z = 0
\,.
\end{equation}

In the paramagnetic phase, the magnetizations and overlaps are zero, and the expression for the free energy Eq.~\ref{eq:free_energy_RS} simplifies to
\begin{equation}
\label{eq:paramagnetic_energy}
\beta [f] = - \frac{\sqrt{\alpha_v \alpha_h}}{2} \bar{\omega}^2 - \log 2 \,.
\end{equation}
This expression was first derived in \cite{barra2011equilibrium}, and it was also proven that this replica-symmetric expression is valid in this phase. In the SK model, the assumption of replica symmetry fails to hold in the spin glass phase and in a low-temperature subset of the ferromagnetic phase due to the presence of unstable modes, as shown by \cite{de1978stability}. We expect the same is true for the bipartite SK model, however we will omit a detailed stability analysis here. Instead, the unphysical nature of the replica symmetric solution may be illustrated by the fact that it predicts a negative entropy in the zero temperature limit within the spin glass phase. In particular, setting $w_0 = 0$, the zero temperature limit of the saddle equations Eq.~\ref{eq:RSsaddle} yield zero magnetization, $m^{(v)} = m^{(h)} = 0$, as well as
\begin{equation}
q^{(v)} = 1 - \sqrt{\frac{2}{\pi}} \left( \frac{\alpha_v}{\alpha_h} \right)^{1/4} \frac{T}{w} + \mathcal{O}(T^2) \,, \qquad v \leftrightarrow h.
\end{equation}
This may be inserted into the free energy to yield
\begin{equation}
\label{eq:fRS_zerotemp}
[f] = - \sqrt{\frac{2 \alpha_v \alpha_h}{\pi}} w \left[ \left(\frac{\alpha_v}{\alpha_h} \right)^{1/4} + \left(\frac{\alpha_h}{\alpha_v} \right)^{1/4} \right] + \frac{\sqrt{\alpha_v \alpha_h}}{\pi} T + \mathcal{O}(T^2) \,.
\end{equation}
As the free energy is related to the entropy via $S = -\partial F / \partial T$ at fixed volume and particle number, the assumption of replica symmetry predicts a negative entropy density ${S/N = -\sqrt{\alpha_v \alpha_h}/\pi}$, which is unphysical.

\subsection{Replica Symmetry Breaking Analysis}
In this section we drop the assumption of replica symmetry, which fails to hold in the spin glass phase. (\cite{oppermann2004modulated} performed a similar calculation for the related Korenblit-Shender model defined in footnote \ref{ftnt:KS}, in which $\alpha_h = 1/2$; here we consider zero biases but arbitrary $\alpha$.)

\subsubsection{Review: Parisi Ansatz \label{sec:parisireview}}
Parisi famously proposed in \cite{parisi1980sequence} a very interesting ansatz for replica symmetry breaking in the context of the original unipartite SK model. We will consider the natural extension of his ansatz to the bipartite model studied here. In order for the presentation to be as self-contained as possible, we first briefly review the Parisi ansatz for the unipartite SK model for a single set of $N$ spins $s_i$, with Hamiltonian given by Eq.~\ref{eq:SKHamiltonian}.

The spin glass phase of the unipartite SK model is characterized by a free energy with many local minima separated by high free energy barriers. As a result, thermodynamic averages may be decomposed into sums of pure states. If the pure states are indexed by $a$ (the reason for using the same index as for the replicas will become apparent soon), then for any observable $\mathcal{O}$, 
\begin{equation}
\langle \mathcal{O} \rangle = \frac{\text{Tr}\left(e^{-\beta H} \mathcal{O}\right)}{Z} = \sum_a w_a \langle \mathcal{O} \rangle_{a} \,,
\end{equation}
where the weights of the pure states are given by $w_a = e^{-F_a}/Z$, with $F_a$ the free energy of the pure state. Pure states satisfy cluster decomposition, which for an infinite-dimensional model such as the bipartite SK model implies that correlation functions of observables at different sites must factorize: $\langle \mathcal{O}_{i_1} \mathcal{O}_{i_2} \rangle_{a} = \langle \mathcal{O}_{i_1} \rangle_{a} \langle \mathcal{O}_{i_2} \rangle_{a}$. The large free energy barriers between pure states cause ergodicity to be broken, so that the system becomes trapped in a single pure state during time evolution.

The relationship between the existence of many distinct pure states and the replica method was first worked out by Parisi \cite{parisi1980sequence}, who showed that the matrix which measures the overlap between replicas $a$ and $b$ may be identified with the matrix which measures the overlap between pure states $a$ and $b$. This corresponds to equating
\begin{equation}
Q_{a b} = \frac{1}{N} \sum_i \langle s_i \rangle_a \langle s_i \rangle_b \,,
\end{equation}
and justifies using the same index for replicas and pure states. Under time evolution, ergodicity is broken, and different replicas can spontaneously equilibrate into different pure states despite having been initialized equivalently. Thus, to properly capture the physics of the many pure states, an ansatz for the overlap matrix $Q_{ab}$ must be specified which breaks the permutation symmetry between replicas. 

Parisi introduced a particularly interesting ansatz, which may be constructed according to the following prescription. First, a series of $k+2$ integers is introduced, $m_I$, $I=0,...,k+1$, with $m_0 = n$, and $m_{k+1} = 1$, and all $m_I/m_{I+1}$ assumed to be integers. The off-diagonal elements of the $n \times n$ overlap matrix are then parametrized according to
\begin{equation}
\label{eq:parisi_ansatz}
Q_{ab} = q_I \,, \quad \text{for} \quad \left\lceil \frac{a}{m_I} \right\rceil = \left\lceil\frac{b}{m_I} \right\rceil \quad \text{and} \quad \left\lceil\frac{a}{m_{I+1}} \right\rceil \neq \left\lceil\frac{b}{m_{I+1}} \right\rceil \,,
\end{equation}
where $\lceil x \rceil$ is the ceiling function. The $q_I$ form a sequence of variational parameters which replace the role of $q$ in the replica-symmetric analysis. An example of this ansatz is depicted below in Fig.~\ref{fig:parisi_ansatz}.
\begin{figure}[h!]
\centering
\includegraphics[width=0.4\textwidth]{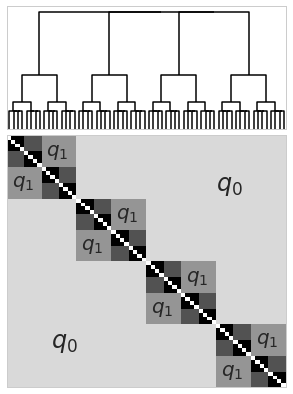}   
\caption{\label{fig:parisi_ansatz}The block-diagonal Parisi ansatz for the overlap matrices $Q_{a b}$ (bottom) along with a dendrogram (top) illustrating the hierarchical relationship contained within the ansatz. Here, $k=3$, with $m_i = (64, 16, 8, 4, 1)$ for $i=0, ..., k+1$. The branching ratio of the tree at level $i$ equals $m_i/m_{i+1}$. This plot uses a grey scale with white corresponding to 0 and black corresponding to 1. The shading thus reflects the fact that the $q_i$ form an increasing sequence (for readability, only $q_0$ and $q_1$ have been labeled).}
\end{figure}

This ansatz breaks the permutation symmetry among replicas in a sequential, iterative manner. The symmetry is broken between pure states belonging to different blocks, and as $k$ increases the number of different blocks increases. The case $k=0$ corresponds to the replica symmetric case considered previously. Finite $k$ corresponds to partial symmetry breaking, and is referred to as the RSB-k scheme. In the extreme case where $k\rightarrow \infty$, the symmetry is fully broken. Regardless of the degree to which replica symmetry is broken, eventually the $n \rightarrow 0$ limit is taken and the overlap matrix $Q_{a b}$ becomes $0\times 0$.

At this point, the relationship between the replica symmetry breaking ansatz for $Q_{ab}$ and the many pure states can be seen by computing
\begin{equation}
P_J(q) = \sum_{a, b} w_{a} w_{b} \delta(q_{a b} - q) \,,
\end{equation}
which is the probability distribution of finding two pure states, $a$ and $b$ with a given overlap $q_{a b} = q$. The $J$ subscript indicates that this expression is not self-averaging; unlike the free energy, it depends on the particular realization for the coupling matrix. After imposing the Parisi ansatz for $k\rightarrow \infty$, taking the disorder average, and the $n\rightarrow 0$ limit, the overlap probability can be shown to be expressible as
\begin{equation}
\label{eq:overlap_to_dist}
P(q) = [ P_J(q)] = \frac{ \mathrm{d} x(q)}{\mathrm{d} q} \,, \qquad x(q) = \int_0^{q} \mathrm{d} q' P(q') \,.
\end{equation}

With $k \rightarrow \infty$, the sequence of overlaps $q_I$ has been converted to a continuous function of a single variable, $q(x)$, and moreover that function has been related to the probability of finding two states with a given overlap. The inverse function $x(q)$ is the cumulative distribution function for the random variable $P$, and gives the probability of finding an overlap less than or equal to $q$. The overlap $q(x)$ is the quantile function for $P$, with domain $x \in [0,1]$ and range $q \in [0, q_{\text{max}}]$.\footnote{\label{ftnt:Z2footnote}A subtley here is the fact that in the zero-bias case, for every pure state there is a $\mathbb{Z}_2$ partner state with the each spin reversed. This means that the overlap functions should lie within the interval $-1 \le -|q_{\text{max}}| \le q(x) \le |q_{\text{max}}| < 1$. The replica symmetry breaking ansatz implicitly assumes that $q(x) \ge 0$, and so the probability distribution across the full interval is related to those of Eq.~\ref{eq:overlap_to_dist} via ${P^{\text{full}}(q) = P(q)/2 + P(-q)/2}$. See p. 32 of \cite{mezard1987spin} for more details on this point.} Thus, breaking of replica symmetry has allowed for a multitude of distinct pure states with different overlaps between each other. 

Perhaps the most fascinating feature of Parisi's replica symmetry breaking solution of the SK model is the fact that the space of pure states is ultrametric \cite{mezard1984nature, mezard1987replica}. Just as $P(q)$ is defined to be the disorder-averaged probability that two arbitrary pure states, $a$ and $b$, will have an overlap $q_{a b} = q$, a 3-state overlap distribution may also be defined. For pure states $a, b, c$, $P(q_1, q_2, q_3)$ may be defined to be the probability that $q_{a b} = q_1$, $q_{b c} = q_2$, and $q_{a c} = q_3$. Interestingly, unlike the case of $P(q)$ which can only be analytically calculated near criticality, the 3-state overlap distribution of the SK model may be calculated in the spin glass phase for any value of the temperature  \cite{mezard1984nature, mezard1987replica}:
\begin{align}
\label{eq:3overlap_dist}
P(q_1, q_2, q_3) =&\ \frac{1}{2} P(q_1) x(q_1) \delta(q_1 - q_2) \delta(q_1 - q_3) \\
&+ \frac{1}{2} \Big( P(q_1) P(q_2) \theta(q_1 - q_2) \delta(q_2 - q_3) + \text{cyclic permutations} \Big) \nonumber \,,
\end{align}
where $\theta(x)$ is the Heaviside theta function.

The simple expression for the 3-overlap distribution, Eq.~\ref{eq:3overlap_dist}, implies that the pure states of the model are arranged in a tree-like structure. This can be best seen by first introducing the distance between pure states:
\begin{equation}
(d_{a b})^2 = \frac{1}{N} \sum_i \left( \langle s_i \rangle_a - \langle s_i \rangle_b \right)^2 = 2\left(q_{EA} - q_{ab} \right) \,,
\end{equation}
where $q_{EA}=q_{aa}=q_{bb}$ is the state-independent Edwards-Anderson order parameter. Assuming without loss of generality that $q_{a b} \le q_{b c} \le q_{a c}$, Eq.~\ref{eq:3overlap_dist} implies that $P(q_1, q_2, q_3)$ is non-zero only for
\begin{equation}
\label{eq:ultrametric_d}
d_{a c} \le d_{a b} = d_{b c} \,.
\end{equation}
Thus, all triangles in the space of states are either equilateral or isosceles with the unequal side shorter than the equal sides.

Metric spaces satisfying Eq.~\ref{eq:ultrametric_d} are called ultrametric.\footnote{The defining equation for ultrametric spaces is normally written as $d_{a c} \le \text{max}\{ d_{a b}, d_{b c} \}$, which is equivalent to Eq.~\ref{eq:ultrametric_d}. For a review of ultrametricity from the perspective of physics, see \cite{rammal1986ultrametricity}.} One interesting property of ultrametric spaces which is relevant here is that points in such a space may be associated with the leaves of a tree diagram, and therefore, the dendrogram in Fig.~\ref{fig:parisi_ansatz} serves to illustrate how the assumption of ultrametricity is implicitly contained within the Parisi ansatz, Eq.~\ref{eq:parisi_ansatz}. The phenomenon of ultrametricity may be understood dynamically. As the temperature is lowered, ancestor pure states split into descendant pure states according to a stochastic branching process \cite{mezard1987spin} depicted in Fig.~\ref{fig:ultrametric_hierarchy} below.

\begin{figure}[h!]
\centering
\includegraphics[width=0.5\textwidth]{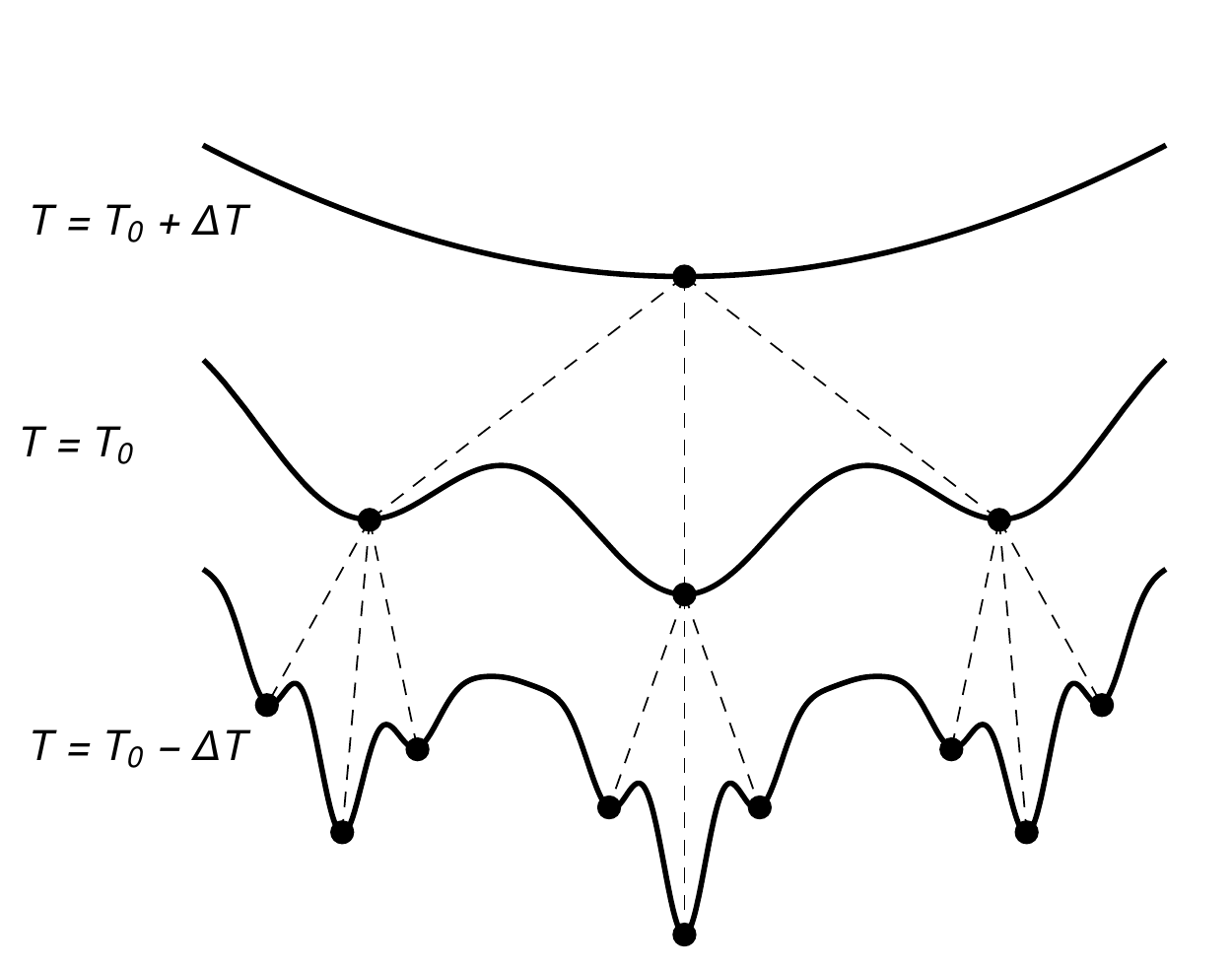}    
\caption{\label{fig:ultrametric_hierarchy}  Schematic depiction of the hierarchical tree of spin glass states, with ancestor states at higher temperatures related to descendant states at lower temperatures by a branching process. The various curves represent the free energy at different temperatures, plotted against a one-dimensional line through the state space. Although the state space is discrete, it is schematically shown here as a continuum.}
\end{figure}

\subsubsection{Free Energy \label{sec:free_energy}}
Given the form of Eq.~\ref{eq:free_energy_no_simplification}, it is natural to separately assume the Parisi ansatz for each of the two overlap matrices, $Q_{ab}^{(v)}$, $Q_{ab}^{(h)}$. We will not attempt to rigorously prove the correctness of this ansatz in the bipartite model, but in Sec.~\ref{sec:mcmc} we show good agreement with Monte Carlo simulations. One immediate implication of the ansatz is that the pure states in the bipartite model will also be ultrametrically distributed. The derivation of Eq.~\ref{eq:3overlap_dist} only relies on the properties of the algebra of Parisi matrices (matrices of the form Eq.~\ref{eq:parisi_ansatz}), and not on the functional form of the Hamiltonian. Thus, by repeating the original calculation for the marginal distributions $p(\bm{v})$, $p(\bm{h})$, one finds that the pure states in each reduced system will exhibit the same ultrametric structure.

The calculation of the free energy proceeds analogously to the SK calculation, which can be found in many textbooks \cite{mezard1987spin, dotsenko2005introduction, nishimori2001statistical}, and so we will not repeat the derivation here. A set of Parisi equations may be derived for infinite breaking of replica symmetry:
\begin{align}
\label{eq:RSB_free_energy_full}
\beta [f] =& -\frac{\sqrt{\alpha_v \alpha_h}}{2} \bar{w}^2 \left(1 + \int_0^1 \mathrm{d} x \, q^{(v)}(x) q^{(h)}(x)  - q^{(v)}(1) - q^{(h)}(1) \right) \\
& - \alpha_v \left\langle f_0^{(h)}(0, \sqrt{r^{(h)}(0)} z) \right\rangle_z - \alpha_h \left\langle f_0^{(v)}(0, \sqrt{r^{(v)}(0)} z) \right\rangle_z \,, \nonumber
\end{align}
where we have introduced
\begin{equation}
\label{eq:rdef}
r^{(v)}(x) = \sqrt{\frac{\alpha_v}{\alpha_h}} q^{(v)}(x) \, , \qquad v \leftrightarrow h
\end{equation}
to simplify the expressions. The functions $f_0^{(I)}(x,h)$, for $I \in \{v, h\}$, are determined to be a solution of the differential equation:
\begin{equation}
\frac{\partial f^{(I)}_0}{\partial x} = - \frac{w^2}{2} \frac{\mathrm{d} r^{(I)}}{\mathrm{d} x} \left( \frac{\partial^2 f_0^{(I)}}{\partial h^2} + x \left( \frac{\partial f_0^{(I)}}{\partial h} \right)^2 \right) \,,
\end{equation}
The boundary condition is that $f_0^{(I)}(1,h) = \log 2 \cosh \beta h$. 

These equations may be solved to a high degree of numerical precision using the methods of \cite{crisanti2002analysis}. As these techniques are somewhat involved, we will instead be content to solve the equations in the RSB-1 scheme (see Sec.~\ref{sec:parisireview}). The $n \to 0$ limit converts the integer $m_1$ into a variational parameter in the interval $(0, 1)$ which we denote by $m$ (which should not be confused with the magnetization). The other variational parameters are a sequence of $q$-values for each spin species: $\left\{q_0^{(v)}, q_1^{(v)}\right\}$, $\left\{q_0^{(h)}, q_1^{(h)}\right\}$. The RSB-1 expression for the free energy is
\begin{align}
\label{eq:RSB1_free_energy}
\beta [f] =& -\frac{\sqrt{\alpha_v \alpha_h}}{2} \bar{w}^2 \left( 1 + m q_0^{(v)} q_0^{(h)} + (1-m) q_1^{(v)} q_1^{(h)} - q_1^{(v)} - q_1^{(h)} \right) - \log 2 \\
&- \frac{\alpha_v}{m} \int_{-\infty}^{\infty} \mathrm{d}P\left(r_0^{(h)},z\right)\log \left( \int_{-\infty}^{\infty} \mathrm{d} P\left(r_1^{(h)} - r_0^{(h)}, y\right) \cosh^m \left[ \beta(y + z) \right] \right) \nonumber \\
&- \frac{\alpha_h}{m} \int_{-\infty}^{\infty} \mathrm{d}P\left(r_0^{(v)},z\right) \log \left( \int_{-\infty}^{\infty} \mathrm{d} P\left(r_1^{(v)} - r_0^{(v)}, y\right) \cosh^m \left[ \beta(y + z) \right] \right) \,, \nonumber
\end{align}
where $\mathrm{d}P\left(q,z\right) := \frac{\mathrm{d} z}{\sqrt{2\pi q}} \exp\left( - \frac{z^2}{2 q} \right)$ is a Gaussian measure, and the $r$ variables are related to the $q$ variables as in Eq.~\ref{eq:rdef}. The variational parameters are set to the critical point at which ${\bf \nabla}[f] = {\bf 0}$, subject to the constraint that $0 < m < 1$, ${0 < q_0^{(v)} < q_1^{(v)} < 1}$, and similarly for $q_0^{(h)}, q_1^{(h)}$. We present the numerical solution of this equation, along with the results of MC simulations, in Sec.~\ref{sec:mcmc} below.\footnote{\label{ftnt:RSB1footnote}One new subtlety that arises in the bipartite case is that the critical point turns out to be a saddle point of $[f]$, instead of a local maximum as in the unipartite case. We have two possible explanations for this phenomenon: (1) Even in the unipartite case, the local extremum of $[f]$ within the RSB-$k$ ansatz at finite $k$ is known to become a saddle point if we allow perturbations that violate the ansatz \cite{de1978stability}. The more general bipartite RSB-$k$ ansatz may directly include these ``unstable directions'' in parameter space as an artifact of the finite value of $k$. (2) The integration contour deformation implicit in the change of variables Eq.~\ref{eq:complexvariables} may not lie along the direction of steepest descent of $\left| e^{-\beta N n \mathcal{F}_n} \right|$. In this case a more careful integration contour deformation might convert the free energy's critical point from a saddle point to a local extremum. However, the orientation of the integration contour in the complex plane as it passes through the critical point only makes a contribution subleading in large $N$, and so is irrelevant in the $N \to \infty$ limit. Regardless of the explanation, the saddle nature of the critical point makes it much more difficult to locate numerically than a local extremum, especially at low temperature, as we must simultaneously solve the five equations ${\bf \nabla}[f] = {\bf 0}$ instead of simply extremizing a single scalar quantity.}

\subsubsection{Near-Critical Solution}
Analytic results for the spin glass phase are in general not possible for the replica symmetry breaking analysis. However, they are obtainable in the vicinity of the spin-glass phase transition. Here we will work out the near-critical overlap distributions, and verify that in the RSB analysis the bipartite SK model exhibits a spin glass transition which is very similar to that of the standard, unipartite model. For simplicity, we will again set $\bar{w}_0 = 0$.

The starting point is the free energy, Eq.~\ref{eq:free_energy_no_simplification}. In the paramagnetic phase, the overlap matrices will be zero, and in the spin glass phase, they will be non-zero. In the spin glass phase and near the phase boundary $\bar{w} = 1$, they will be non-zero but small, and therefore the free energy may be expanded as
\begin{align}
\beta n \mathcal{F}_n = &\ \frac{\sqrt{\alpha_v \alpha_h} \bar{w}^2}{2} \left( \text{Tr} \left[Q^{(h)} Q^{(v)}\right]  - n \right) - n \log 2 \\
& - \alpha_v \Bigg[ \left(\frac{\alpha_h}{\alpha_v}\right) \frac{\bar{w}^4}{4} \text{Tr} \left[\left(Q^{(h)}\right)^2\right] + \left(\frac{\alpha_h}{\alpha_v}\right)^{3/2} \frac{\bar{w}^6}{6} \text{Tr} \left[\left(Q^{(h)}\right)^3\right] + \left(\frac{\alpha_h}{\alpha_v}\right)^2 \frac{\bar{w}^8}{12} \sum_{a b} \left[\left(Q_{ab}^{(h)}\right)^4\right] \nonumber \\
& \qquad \qquad - \left(\frac{\alpha_h}{\alpha_v}\right)^2 \frac{\bar{w}^8}{4} \sum_{a b c} \left[\left(Q_{a b}^{(h)}\right)^2 \left(Q_{a c}^{(h)}\right)^2\right] + \left(\frac{\alpha_h}{\alpha_v}\right)^2 \frac{\bar{w}^8}{8} \text{Tr} \left[\left(Q^{(h)}\right)^4\right] \Bigg] \nonumber \\
& - \alpha_h \Bigg[ \left(\frac{\alpha_v}{\alpha_h}\right) \frac{\bar{w}^4}{4} \text{Tr} \left[\left(Q^{(v)}\right)^2\right] + \left(\frac{\alpha_v}{\alpha_h}\right)^{3/2} \frac{\bar{w}^6}{6} \text{Tr} \left[\left(Q^{(v)}\right)^3\right] + \left(\frac{\alpha_v}{\alpha_h}\right)^2 \frac{\bar{w}^8}{12} \sum_{a b} \left[\left(Q_{ab}^{(v)}\right)^4\right] \nonumber \\
& \qquad \qquad - \left(\frac{\alpha_v}{\alpha_h}\right)^2 \frac{\bar{w}^8}{4} \sum_{a b c} \left[\left(Q_{a b}^{(v)}\right)^2 \left(Q_{a c}^{(v)}\right)^2\right] + \left(\frac{\alpha_v}{\alpha_h}\right)^2 \frac{\bar{w}^8}{8} \text{Tr} \left[\left(Q^{(v)}\right)^4\right] \Bigg] \,, \nonumber
\end{align}
where we have carried out the expansion to fourth order and have only kept the quartic terms which contribute to the phase transition at leading order.

To further analyze this expression, an ansatz for the overlap matrices $Q_{ab}^{(v)}$, $Q_{ab}^{(h)}$ must be specified in order to carry out the traces. A natural choice, given the similarity of the bipartite SK model to the unipartite SK model, is to assume that each of these matrices is of the Parisi form \cite{dotsenko2005introduction, mezard1987spin, nishimori2001statistical}. In the limit $n\rightarrow 0$, the matrices may be written in terms of continuous functions $q^{(v)}(x)$, $q^{(h)}(x)$, and the traces may be carried out, yielding
\begin{align}
\beta [f] = \int_0^1 \mathrm{d} x \Bigg[& \frac{1}{4} \left(\sqrt{\alpha_h} q^{(h)} -\sqrt{\alpha_v} q^{(v)} \right)^2 + \sqrt{\alpha_v \alpha_h}  q^{(v)}(x) q^{(h)}(x) \tau \\
& - \frac{\alpha_h^{3/2}}{6\alpha_v^{1/2}} \left( x q^{(h)}(x)^3 + 3 q^{(h)}(x) \int_0^x \mathrm{d}u\, q^{(h)}(u)^2 \right) \nonumber\\
& - \frac{\alpha_v^{3/2}}{6\alpha_h^{1/2}} \left( x q^{(v)}(x)^3 + 3 q^{(v)}(x) \int_0^x \mathrm{d}u\, q^{(v)}(u)^2 \right) \nonumber\\
& + \frac{\alpha_h^2}{12 \alpha_v} q^{(h)}(x)^4 + \frac{\alpha_v^2}{12 \alpha_h} q^{(v)}(x)^4 \Bigg] + C \,. \nonumber
\end{align}
Here we have expanded around the critical point, setting $\bar{w} = 1 + \tau$, with $\tau > 0$ and $\tau \ll 1$. $C$ is an irrelevant overall constant. Additionally, we will neglect the first term; this will be justified by the solution \textit{a posteriori}.

Next, the variational derivative with respect to both overlap function may be taken, yielding
\begin{align}
0 =\, &\sqrt{\alpha_v \alpha_h} q^{(v)}(x) \tau \\
& - \frac{\alpha_h^{3/2}}{6 \alpha_v^{1/2}} \left( 3 x q^{(h)}(x)^2 + 3 \int_0^x \mathrm{d} u \, q^{(h)}(u)^2 + 6 q^{(h)}(x) \int_x^1 \mathrm{d} u \, q^{(h)}(u) \right) + \frac{\alpha_h^2}{3 \alpha_v} q^{(h)}(u)^3 \,, \nonumber \\
v \leftrightarrow &\, h. \nonumber
\end{align}

These equations are solved (to leading order in $\tau$) by
\begin{equation}
\label{eq:nearcriticaloverlaps_appendix}
q^{(v)}(x) = 
   \begin{cases} 
      \sqrt{\frac{\alpha_h}{\alpha_v}} \frac{x}{2} &  0 \le x \le 2 \tau \\
      \sqrt{\frac{\alpha_h}{\alpha_v}} \tau & 2\tau \le x \le 1 
   \end{cases}
   \,, \qquad v \leftrightarrow h.   
\end{equation}

\begin{figure}
\centering
\begin{subfigure}{.5\textwidth}
  \centering
\includegraphics[page=3, width=1.0\textwidth]{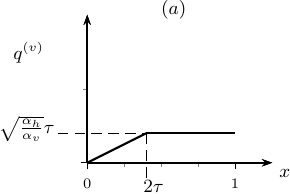}  
\end{subfigure}%
\begin{subfigure}{.5\textwidth}
  \centering
\includegraphics[page=4, width=1.0\textwidth]{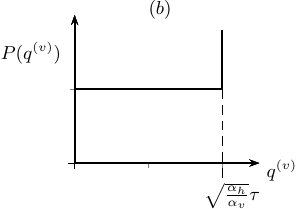}    
\end{subfigure}
\caption{\label{fig:criticaloverlapplots} (a): The near-critical overlap $q^{(v)}$ for the visible spins. The overlap for the hidden spins may be obtained by interchanging $\alpha_v \leftrightarrow \alpha_h$. (b): The probability distribution of visible overlaps. The vertical line represents a delta function spike.}
\end{figure}

Using Eq.~\ref{eq:overlap_to_dist}, the disorder-averaged probability distributions of the overlaps may be found to be non-zero only in the intervals $q^{(v)} \in \left[0, \sqrt{\frac{\alpha_h}{\alpha_v}} \tau\right]$, $v \leftrightarrow h$, in which case they are
\begin{equation}
\label{eq:critical_overlap_dist}
P\left(q^{(v)}\right) = \left(1 - 2\tau \right) \delta\left( q^{(v)} - \sqrt{\frac{\alpha_h}{\alpha_v}} \tau \right) + 2 \sqrt{\frac{\alpha_v}{\alpha_h}} \,, \qquad v \leftrightarrow h.
\end{equation}
The Dirac delta function terms in Eq.~\ref{eq:critical_overlap_dist} correspond a non-zero probability $(1-2\tau)$ that two random pure states will have an overlap  equal to the largest value in the interval. The constant 2 corresponds a non-zero probability for the overlap to take on a value within the interval. The near-critical overlaps and their distributions are plotted in Fig.~\ref{fig:criticaloverlapplots}.

In Section~\ref{sec:mcmc} the analytic predictions of the RSB analysis will be compared with Monte Carlo (MC) simulations, and the two methods will be found to be in good agreement.

\section{Application: Graph Partitioning \label{sec:optimization}}
In this section, we consider an application of the bipartite SK model to a combinatorial optimization problem. By solving this combinatorial problem we will obtain an independent check on the RSB and MC results. 

There is a rich connection between the replica method and optimization problems; for example, in \cite{mezard1987replica} it was used to study the traveling salesman problem. Of course, the replica method cannot yield a solution to a general instance of the problem (which would be extremely surprising), but can instead allow for various ``thermodynamic'' quantities to be calculated, such as the average length of the shortest path. In another application of spin glass theory to optimization problems, Fu and Anderson \cite{fu1987application} considered an NP-complete graph partitioning problem and related it to the SK model. Given a graph $G(V,E)$ with vertex and edge sets $V,E$ respectively, the problem is to find the partitioning of the vertices into two equally-sized subsets, such that the number of edges between the two subsets is minimized.\footnote{This problem corresponds to ND14 in Garey and Johnson \cite{garey2002computers}.}

Fu and Anderson considered a special case of this problem, and took the graph to be an instance of a random graph for which each edge is present with probability $p$. Defining $N:=|V|$, they considered the large-$N$ limit with $p$ kept finite, so that each vertex has an infinite number of connections. They showed that the cost function $C$, when evaluated on the optimal solution, grows as
\begin{equation}
C^* \sim \frac{N^2}{4}p + N^{3/2} \sqrt{p (1-p) } \frac{E^{\text{SK}}_{gs}}{2 N} \,.
\end{equation}
$C$ measures the number of edges between the partitions and the $*$ indicates that the cost has been evaluated on the optimal solution. Here $E^{\text{SK}}_{gs}/N$ is the ground-state energy density of the original SK model.\footnote{Here it has been assumed that $J_{ij} \sim \mathcal{N}(0, N^{-1/2})$.} The energy can be calculated via the replica method to be ${E^{\text{SK}}_{gs}/N = -0.7633}$ \cite{mezard1987spin}. Note that the term involving the energy appears as a subleading correction to the optimal cost.

\begin{figure}[!ht]
\centering
\includegraphics[page=6, width=0.6\textwidth]{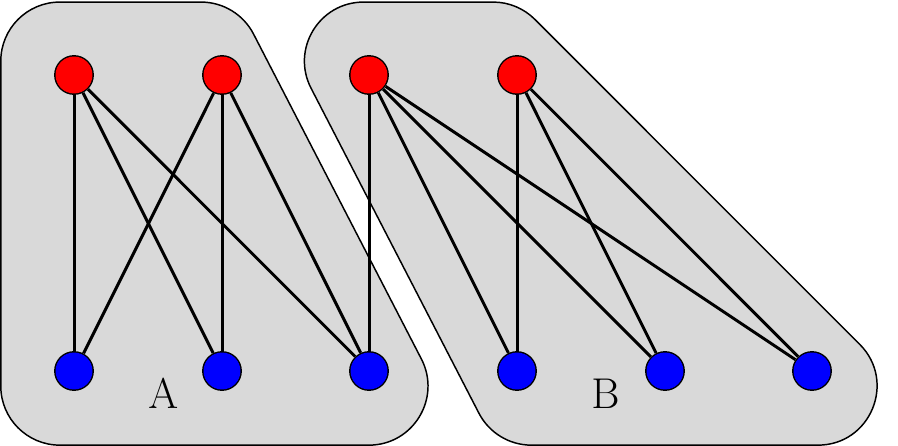}
\caption{\label{fig:graph_partitioning} A depiction of the bipartite graph partitioning problem. The vertices have been separated into two clusters, $A, B$, with each cluster containing $N_v/2$ vertices of one species, and $N_h/2$ of the other. The clusters were chosen to minimize the number of inter-partition edges.}
\end{figure}

A bipartite extension of the graph partitioning problem may be formulated, which can then be related to the ground state of the bipartite SK model. The problem is: given a bipartite graph with $N_v$ vertices of one type and $N_h$ of another (with both $N_v$ and $N_h$ assumed to be even), find a clustering of the vertices into two equally-sized subsets each with $N_v/2$ vertices of one part and $N_h/2$ vertices of the other part, such that the number of edges between vertices in separate clusters is minimized.\footnote{One possible source of confusion here is that there are two different notions of partitions - the partitioning implied by the bipartite structure, and the partitioning which is the goal of the optimization problem. We will refer to this second partitioning as a ``cluster'' to avoid confusion.} This is depicted in Fig.~\ref{fig:graph_partitioning}.

The version of the bipartite graph partitioning problem that we study here is the following. Of the $N_v N_h$ possible edges, each one is present independently with probability $p$. With minor modifications, the treatment of Fu and Anderson \cite{fu1987application} can be applied to this problem as well. Let $i=1,...,N_v$ label the vertices in the first partition, and $j=1,...,N_h$ label the vertices in the second. Also let $\Omega_{ij}$ be a matrix used to indicate whether a given edge is present, with $\Omega_{ij} = \Omega$ if the edge between vertices $(i,j)$ is present, and 0 otherwise. Let $A$ and $B$ be the two clusters, with $v_i, h_j = +1$ if the $i$-th or $j$-th vertex is assigned to cluster $A$, and $v_i, h_j = - 1$ if it is assigned to cluster $B$.

In order to relate this problem to the bipartite SK model, we introduce the Hamiltonian $H = - \sum_{i,j} \Omega_{ij} v_i h_j$. Evaluated on any spin configuration, the Hamiltonian may be related to the cost function $C$ as
\begin{equation}
\label{eq:costeqn}
C = \frac{N_v N_h}{2} p + \frac{H}{2 \Omega} \,.  
\end{equation}
$C$ again counts the number of edges between clusters $A$ and $B$. The cost of the optimal solution can then be related to the minimal value of the Hamiltonian -- in other words the ground state energy:
\begin{equation}
C^* = \frac{N_v N_h}{2} p + \frac{E_{gs}}{2 \Omega} \,. 
\end{equation}

Since the energy is a self-averaging quantity, the ground state energy of the Hamiltonian $H$ may be computed by taking the disorder average and applying the replica method. The calculation is similar to that of the bipartite SK model, although it is not identical because of the important difference that the couplings $\Omega_{ij}$ are now Bernoulli random variables rather than Gaussian, and there is an additional constraint that each cluster should contain half of the visible and half of the hidden vertices, so that ${\sum_i v_i = \sum_j h_j = 0}$. Nevertheless, the ground state energies of the two Hamiltonians are the same in the large-$N$ limit, up to an overall scaling, as we will now show.

The calculation proceeds by first computing the free energy $[F]$ for arbitrary temperature. Then, the zero-temperature limit may be taken, in which case the free-energy reduces to the ground-state energy. As above, to compute $[F]$ we first compute $[Z^n]$,
\begin{align}
[Z^n] &= \left(1-p\right)^{N_v N_h} \text{Tr}' \prod_{i,j} \left[ 1 + \frac{p}{1-p} \exp\left( \beta \Omega \sum_{a} v_i^{a} h_j^{a} \right) \right] \\
&= \left(1-p\right)^{N_v N_h} \text{Tr}' \exp\left[ \sum_{ij} \log \left( 1 + \frac{p}{1-p} \exp\left( \beta \Omega \sum_{a} v_i^{a} h_j^{a} \right) \right) \right] \,. \nonumber
\end{align}
where $\text{Tr}'$ indicates that the trace is constrained to be over all configurations of the spins such that ${\sum_i v_i^{a} = 0}$, ${\sum_j h_j^{a} = 0}$. By expanding the logarithm and the second exponential, interchanging the summations, and dropping terms which vanish due to the constraints, one finds
\begin{equation}
\label{eq:Zngraphpartition}
[Z^n] = \text{Tr}' \exp\left[ \sum_{\ell = 2}^{\infty} c_{\ell} \left( \beta \Omega \right)^{\ell} N_v N_h \sum_{a_1} ... \sum_{a_{\ell}} \left( \frac{1}{N_v} \sum_i v_i^{a_1} ... v_i^{a_{\ell}} \right) \left( \frac{1}{N_h} \sum_i h_i^{a_1} ... h_i^{a_{\ell}} \right) \right] \,,
\end{equation}
with the constants $c_{\ell}$ defined to be
\begin{equation}
c_{\ell} := \frac{1}{\ell!} \sum_{k=1}^{\infty} \frac{(-1)^{k-1}}{k} \left( \frac{p}{1-p} \right)^{k} k^{\ell} \,.
\end{equation}
In order for the large-$N$ limit to exist, we should again set $\Omega^2 = \omega^2/\sqrt{N_v N_h}$, in which case the only term that is relevant in the sum is the $\ell = 2$ term. 

At this point we observe that Eq.~\ref{eq:Zngraphpartition} defines the same partition function as the bipartite SK model, Eq.~\ref{eq:Zbeforeintegralxform}, except for the minor difference that the coupling has been rescaled (as in Sec.~\ref{sec:bipartiteSK}, the biases and the mean $w_0$ have been set to zero). Using the fact that the ground-state energy is linear in the coupling constant, and that $c_2 = p(1-p)/2$, the optimal cost in the large-$N$ limit grows as
\begin{equation}
\label{eq:KL_bipartite_estimate}
C^* \sim \frac{N^2}{2} \alpha_v \, \alpha_h \, p + N^{3/2} (\alpha_v \alpha_h)^{1/4} \sqrt{p(1-p)}\frac{E^{\text{bSK}}_{gs}}{2N w} \,.
\end{equation}
Finally, the ground-state energy $E^{\text{bSK}}_{gs}$ may be computed numerically, either by Monte Carlo simulation or through solution of the replica symmetry breaking equations (for example Eq.~\ref{eq:RSB_free_energy_full}).\footnote{We have glossed over one important point, which is that although the expressions for the free energies Eq.~\ref{eq:Zngraphpartition} and Eq.~\ref{eq:Zbeforeintegralxform} are the same, the traces in the former expression are restricted to obey the constraints $\sum_i v_i^{a} = 0$, $\sum_i h_j^{a} = 0$. Fu and Anderson showed that in the unipartite case, these constraints are automatically satisfied and do not need to be separately imposed. They also provided a general explanation for this which should hold equally well in the bipartite case studied here. Indeed, the numerical results described below confirm this to be the case.}

We numerically tested the replica prediction Eq.~\ref{eq:KL_bipartite_estimate} by simulating many instances of random bipartite graphs and using a simple variant of the well-known Kernighan-Lin (KL) \cite{kernighan1970efficient} algorithm. The KL algorithm applies to the balanced graph partitioning problem, and performs a greedy heuristic search through the space of all clustering assignments. The constraint that each cluster contains exactly half of the vertices is respected by the algorithm, provided that the initial clustering does too. We developed a simple extension of this algorithm to the bipartite problem considered here that (1) enforces that each of the initial clusters contain half of each vertex type (visible and hidden), and (2) only swaps vertices between clusters if both are of the same type. By applying many passes of the algorithm to a given graph, a good estimate of the optimal cost may be obtained.

\begin{figure}[h!]
\centering
\includegraphics[width=0.65\textwidth]{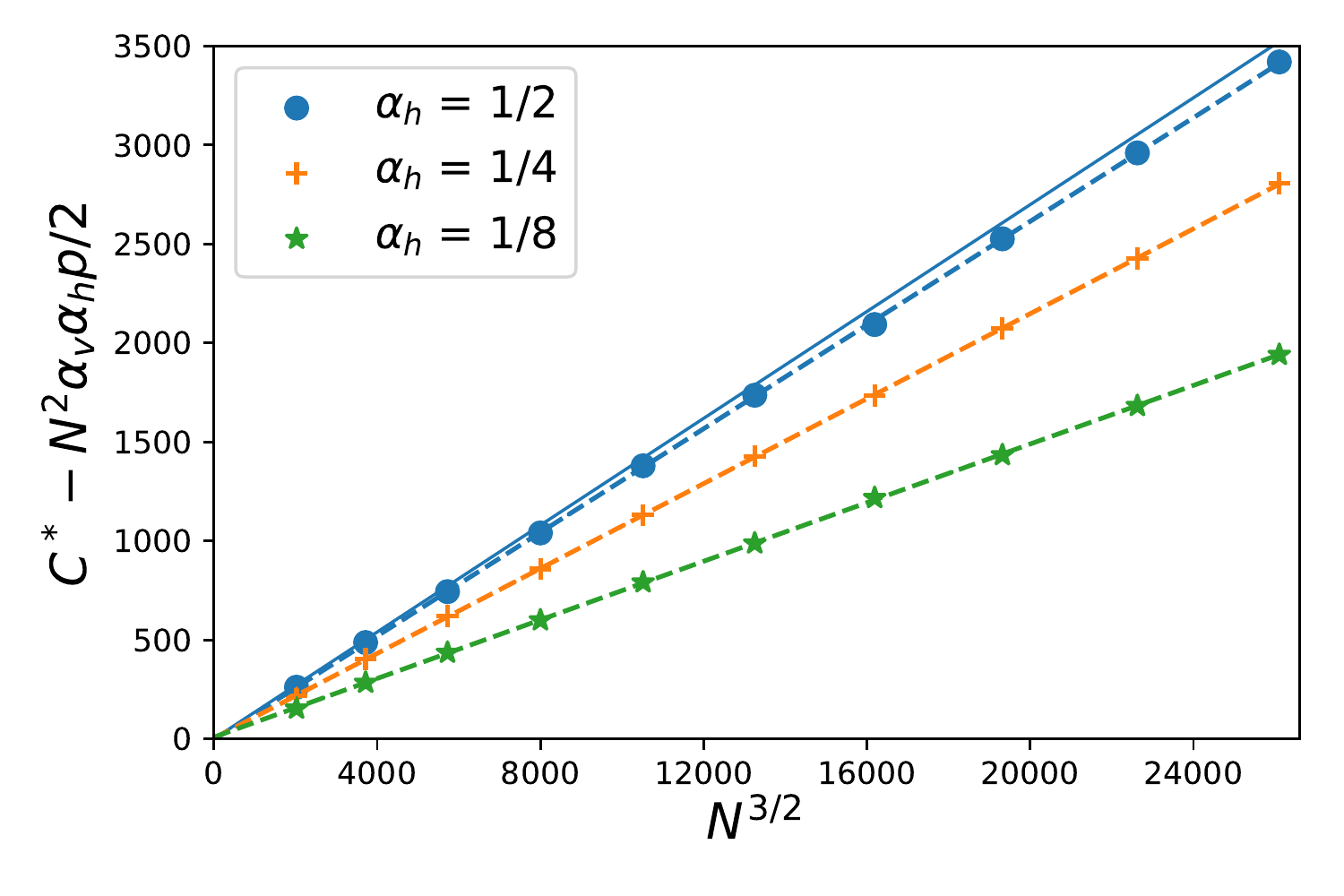}   
\caption{\label{fig:graph_partition_cost}The optimal cost of the graph partitioning problem with the leading term in the large-$N$ limit subtracted, plotted against $N^{3/2}$. The points correspond to simulation results and the dashed lines are linear fits. Eq.~\ref{eq:KL_bipartite_estimate} gives that each line slope is proportional to the ground state energy density of the bipartite SK model with the same value of $\alpha_h$. When $\alpha_h = 1/2$ the ground state energy of the bipartite SK model agrees with that of the unipartite SK model, which is known to high accuracy: $E_{gs}^{\text{SK}} / N = -0.7633$ \cite{mezard1987spin}. The solid line corresponds to the $\alpha_h = 1/2$ analytic prediction, calculated using this value for the ground state energy. The best-fit slope for the simulation data agrees with the analytic prediction to within 2.9\%. In Section \ref{sec:mcmc} below we will compare this method's estimates for $E_{gs}^\text{SK}$ for $\alpha_h \neq 1/2$ against other numerical methods' estimates.}
\end{figure}

In Fig.~\ref{fig:graph_partition_cost} we plot the simulation results for $p = 1/2$ and a range of values for $N$ and $\alpha_h$, and find excellent agreement with the analytical prediction of Eq.~\ref{eq:KL_bipartite_estimate}. The simulation details are as follows: for each $N$, 100 passes of the heuristic KL algorithm were made and the lowest cost was recorded. By repeating this step 100 times and averaging the result, an estimate for $C^*(N)$ was obtained. 

\section{Numerical Results \label{sec:mcmc}}
In this section we present the results of several numerical studies and show that they are in good agreement. First, we numerically solved for the critical point of the RSB-1 free energy Eq.~\ref{eq:RSB1_free_energy}. Then we performed Monte Carlo simulations of the bipartite SK model, using the parallel tempering method \cite{katzgraber2009introduction} to surmount the obstacle of ergodicity breaking. Lastly, we extracted an estimate of the ground state energy by calculating the optimal cost of the graph partitioning problem introduced in the previous section, again using the variant of the Kernighan-Lin algorithm described there. The fact that these rather different methods agree both serves as a check on our calculation and provides strong numerical evidence that the Parisi ansatz is indeed correct for the bipartite SK model.\footnote{Note that the RSB-1 calculation assumes the $N \to \infty$ limit while the MC and graph partitioning simulations necessarily incorporate finite-size effects, so the methods would produce slightly different results even if the numerical simulations were perfectly accurate.}

The MC simulation details are as follows. The bias parameters were again set to zero, as was $w_0$. We also set $w=1$, and worked in terms of the temperature, so that $\bar{w} = 1/T$, and the critical temperature is $T_c = 1$. Using parallel tempering, we simultaneously simulated $N_T = 20$ temperatures for a given disorder realization, with the temperatures distributed geometrically across the interval $[0.1, 4.0]$. We considered $N=400$ spins and 31 different values of $\alpha_h = N_h/N$ uniformly distributed from 0.125 to 0.875.\footnote{The all-to-all nature of the bipartite couplings results in far more bonds than are present in a local system with the same number of spins. This prevents us from reaching system sizes as large as can be reached for local systems, but also means that we do not need to go to nearly as many spins in order to reach the large-$N$ limit.} In order to compute the disorder average, we repeated the experiment for $K = 100$ different random draws of the coupling matrix $W_{ij}$. Each simulation consisted of $10^5$ iterations of a full sweep over all spin variables and a parallel tempering swap. We chose to sample $p(\bm{v}, \bm{h})$, rather than the marginal distributions, because we are interested in both sets of spins. This distribution is also numerically convenient to simulate, because of the fact that in bipartite models all spins within a given partition may be flipped independently of each other due to the factorization property of the conditional probability distribution. During each of the $K$ runs, after each iteration we took a sample, omitting the first $100$ samples for a burn-in period. 

For the graph partitioning problem, we also considered $N = 400$ and the same range of $\alpha_h$ values as in the MC simulations. For each value of $\alpha_h$, we performed 40 KL passes over each of 50 random graphs with $p = 1/2$, and averaged the lowest costs to obtain the final estimate, which was then fit to the analytic form Eq.~\ref{eq:KL_bipartite_estimate}.

\begin{figure}
\centering
\begin{subfigure}{.5\textwidth}
  \centering
\includegraphics[width=1.\textwidth]{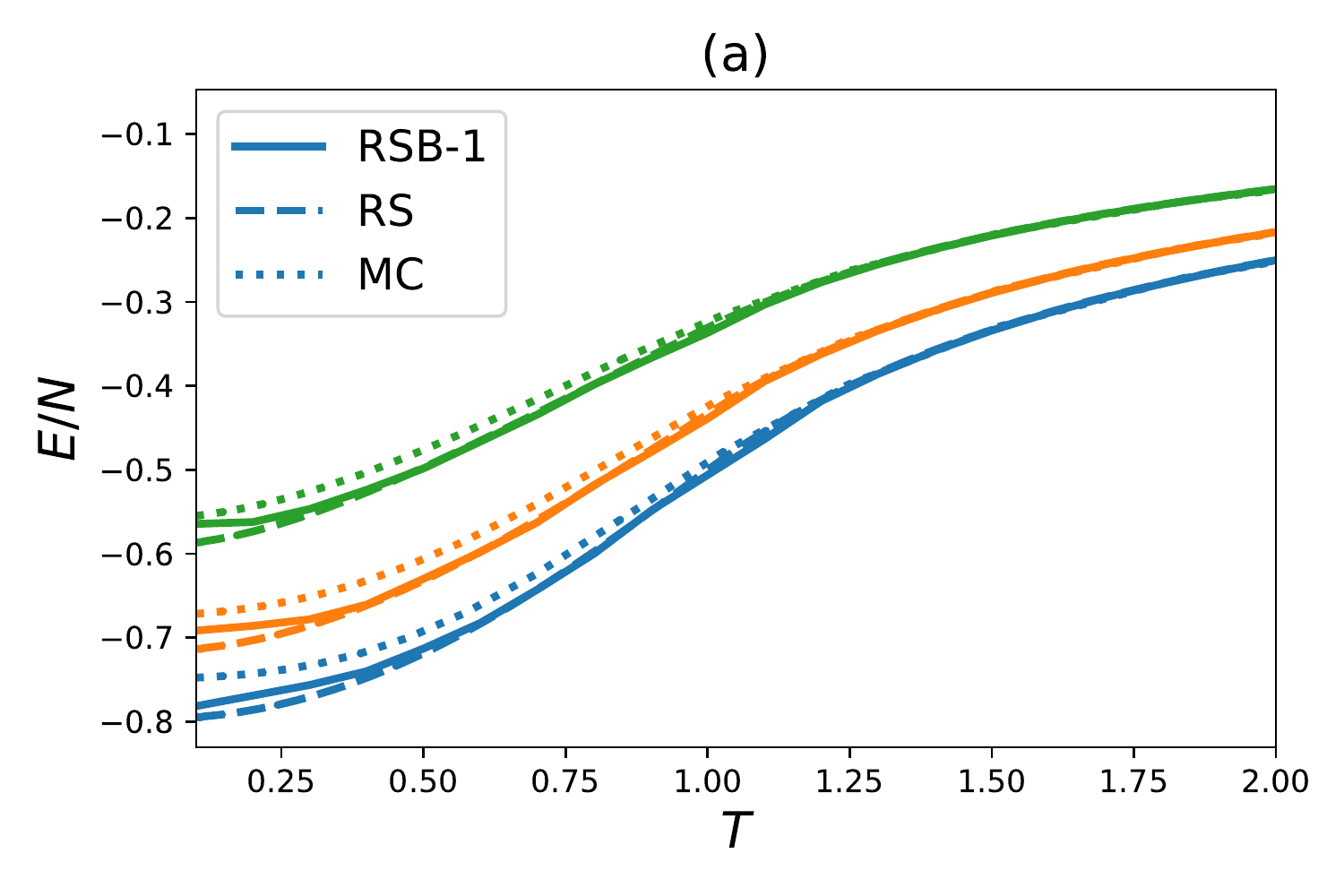} 
\end{subfigure}%
\begin{subfigure}{.5\textwidth}
  \centering
\includegraphics[width=1.\textwidth]{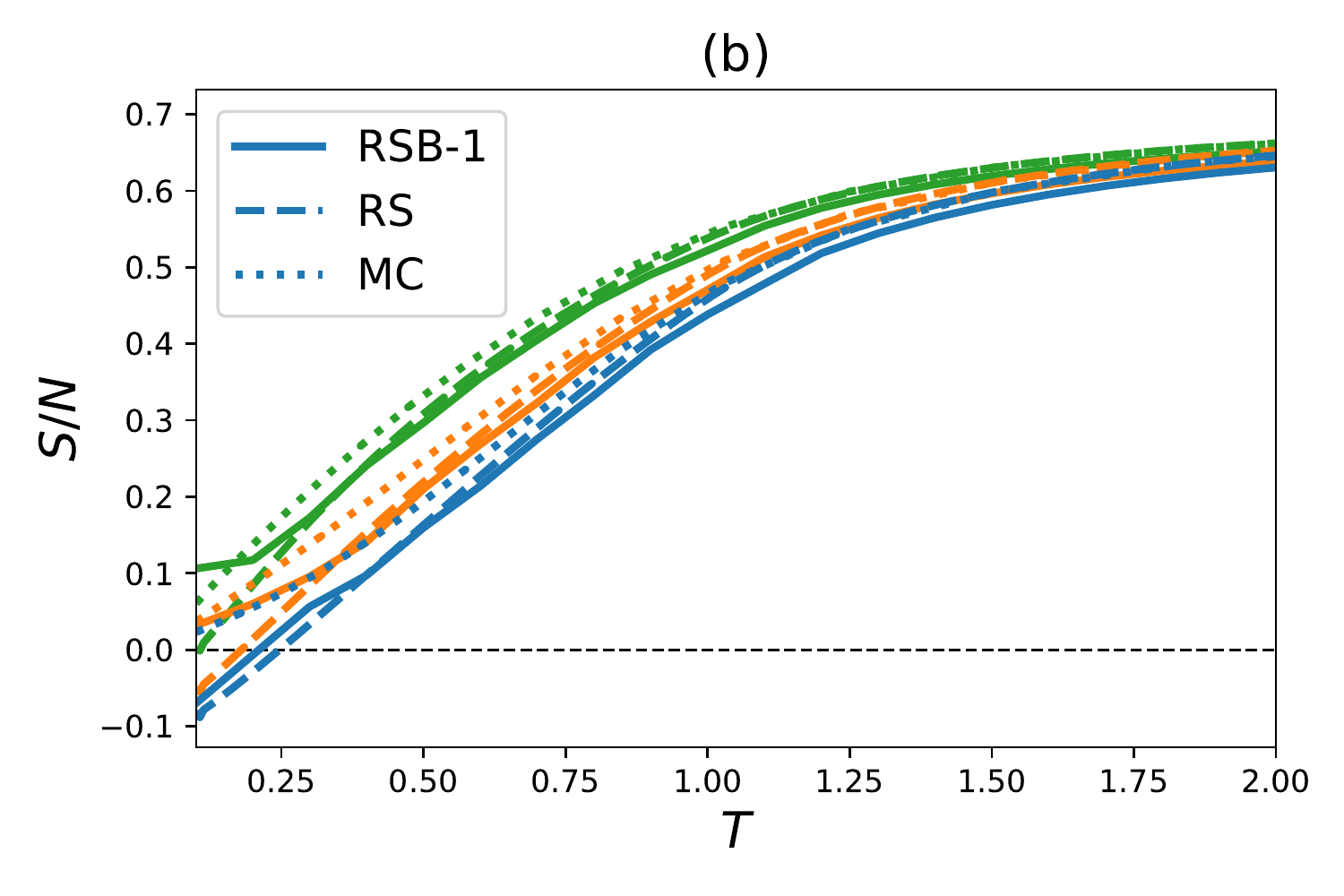}    
\end{subfigure}
\caption{\label{fig:MCenergyentropy} The disorder-averaged energy density (a) and entropy density (b), obtained from 3 distinct calculations and for several values of $\alpha_h$. The bottom set of 3 (blue) curves in both figures correspond to $\alpha_h = 1/2$, the middle set of (orange) curves to $\alpha_h = 1/4$, and the top set of (green) curves to $\alpha_h = 1/8$. Within each set of 3 curves, the solid lines correspond to the RSB-1 ansatz, the dashed to the RS ansatz, and the dotted lines to MC simulations. Above $T>1$, the RS ansatz is exact and gives Eq.~\ref{eq:paramagnetic_energy}, and the RSB-1 result is omitted. In both plots $N = 400$. In (b) the horizontal line corresponds to zero entropy density.}
\end{figure}

We plot the energy and entropy densities as functions of temperature in Fig.~\ref{fig:MCenergyentropy}, from the MC simulations and the RS and RSB-1 approximations. (For $T < 1$ we expect the system to be in the spin-glass phase, so the RSB-1 prediction should be more accurate than the RS prediction.) Overall the agreement is excellent, although at low temperatures there are some differences. The sampling error of the MC simulation increases as the temperature nears 0, and the RSB-$k$ approximation is known to yield a spurious negative entropy for finite $k$ \cite{parisi1980sequence}. Moreover, numerically solving the critical point equations for Eq.~\ref{eq:RSB1_free_energy} in the RSB-1 approximation becomes very challenging at low temperature, as discussed in footnote~\ref{ftnt:RSB1footnote}.

\begin{figure}[h!]
\centering
\includegraphics[width=0.65\textwidth]{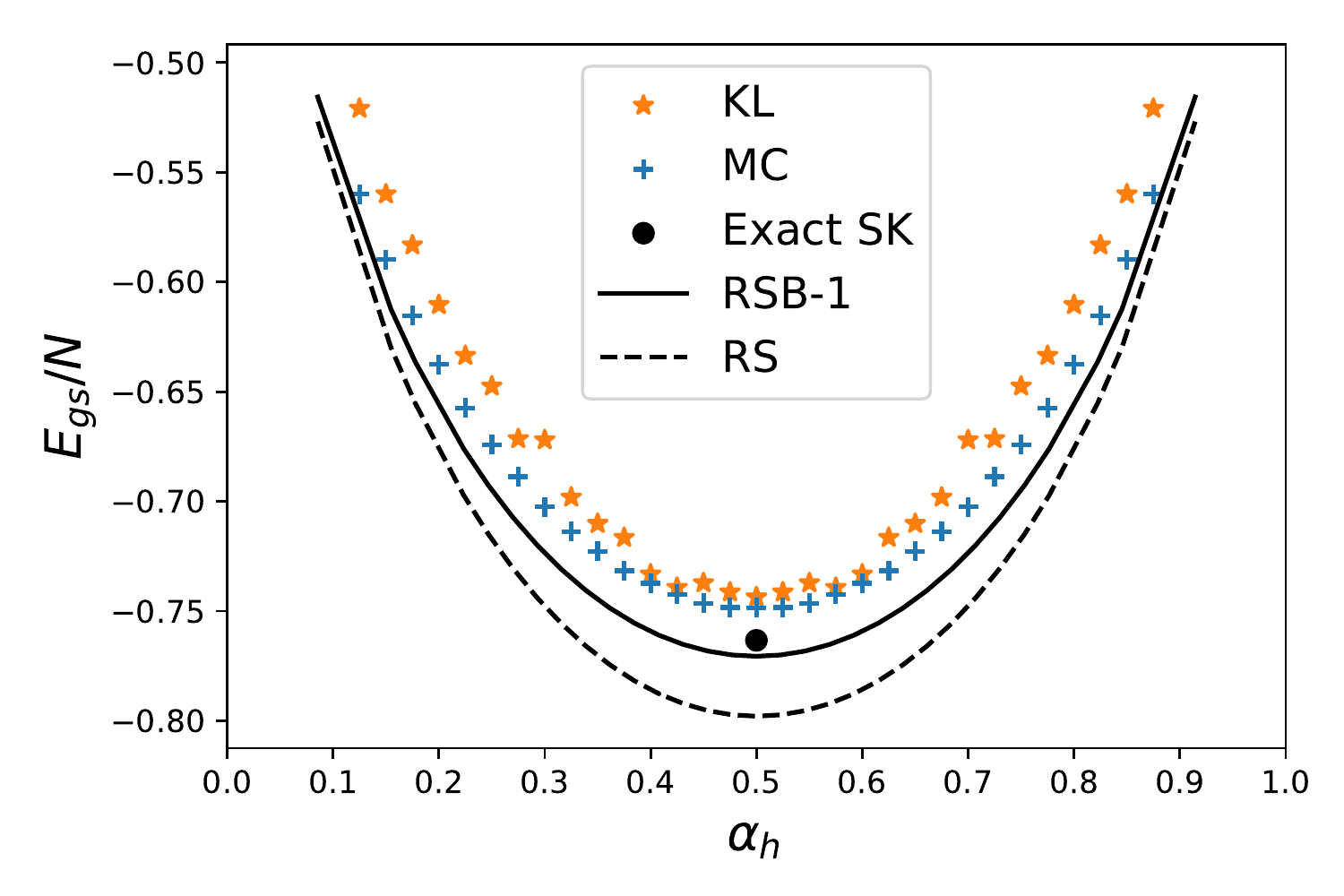}   
\caption{\label{fig:gs_energy} The ground state energy density of the bipartite SK model as a function of the ratio $\alpha_h = N_h/N$. The dashed and solid black curves are the RS and RSB-1 approximations, respectively. The blue crosses correspond to the $T=0$ extrapolation of the Monte Carlo curves shown in Fig.~\ref{fig:MCenergyentropy}. The orange stars correspond to the estimate yielded by the graph partitioning problem considered in Sec.~\ref{sec:optimization} for $N = 400$. The isolated black point at $\alpha_h = 1/2$ corresponds to the exact RSB-$\infty$ result for the ground state energy of the SK model, $E_{gs}^\text{SK}/N = -0.7633$ \cite{mezard1987spin}. For $\alpha_h = 1/2$, the exact ground state energy lies between the RSB-1 and MC estimates, and as explained in the main text, we believe this holds true for general values of $\alpha_h$.}
\end{figure}

The ground state energy as a function of $\alpha_h := N_h/N$ is shown in Fig.~\ref{fig:gs_energy}, where we have also included the predictions from the graph-partitioning problem and the RS symmetric ansatz. We again find good agreement between the methods, although as expected the RS ansatz is significantly less accurate than the RSB-1 ansatz. On general principles, we expect Monte Carlo simulations and the KL algorithm to systematically overestimate the ground state energy of a frustrated system. In the unipartite SK model, the RSB-$k$ approximation systematically \emph{under}estimates the ground state energy by an amount that decreases with $k$ \cite{parisi1980sequence}. Assuming this also holds true in the bipartite case, we therefore expect the exact ground state energy to lie above the RSB-1 prediction but below the MC prediction. 

This is indeed the case for $\alpha_h = \alpha_v = 1/2$, when the result should agree with the original SK model for which $E_{gs}^\text{SK}/N = -0.7633$ \cite{mezard1987spin}: the errors of the various approximations for the ground state energy are RS: $-4.5$\%, RSB-1: $-0.95$\%, MC: $+1.9$\%, KL graph partitioning: $+2.6$\%. That such widely differing approaches yield results which agree to within a few percent is a strong check on our calculations and numerical implementations. It is particularly impressive that the graph partitioning results agree so well with the RSB-1 and MC results, since the graph partitioning estimate for the ground state energy was obtained from an optimal cost $C^{*}$ which was itself only approximately estimated by repeated application of a heuristic algorithm. Remarkably, including only a single level of replica symmetry breaking is enough to improve the replica ansatz's estimate from the least accurate to the most accurate of all the methods we considered.

\begin{figure}
\centering
\begin{subfigure}{.5\textwidth}
  \centering
\includegraphics[width=1.0\textwidth]{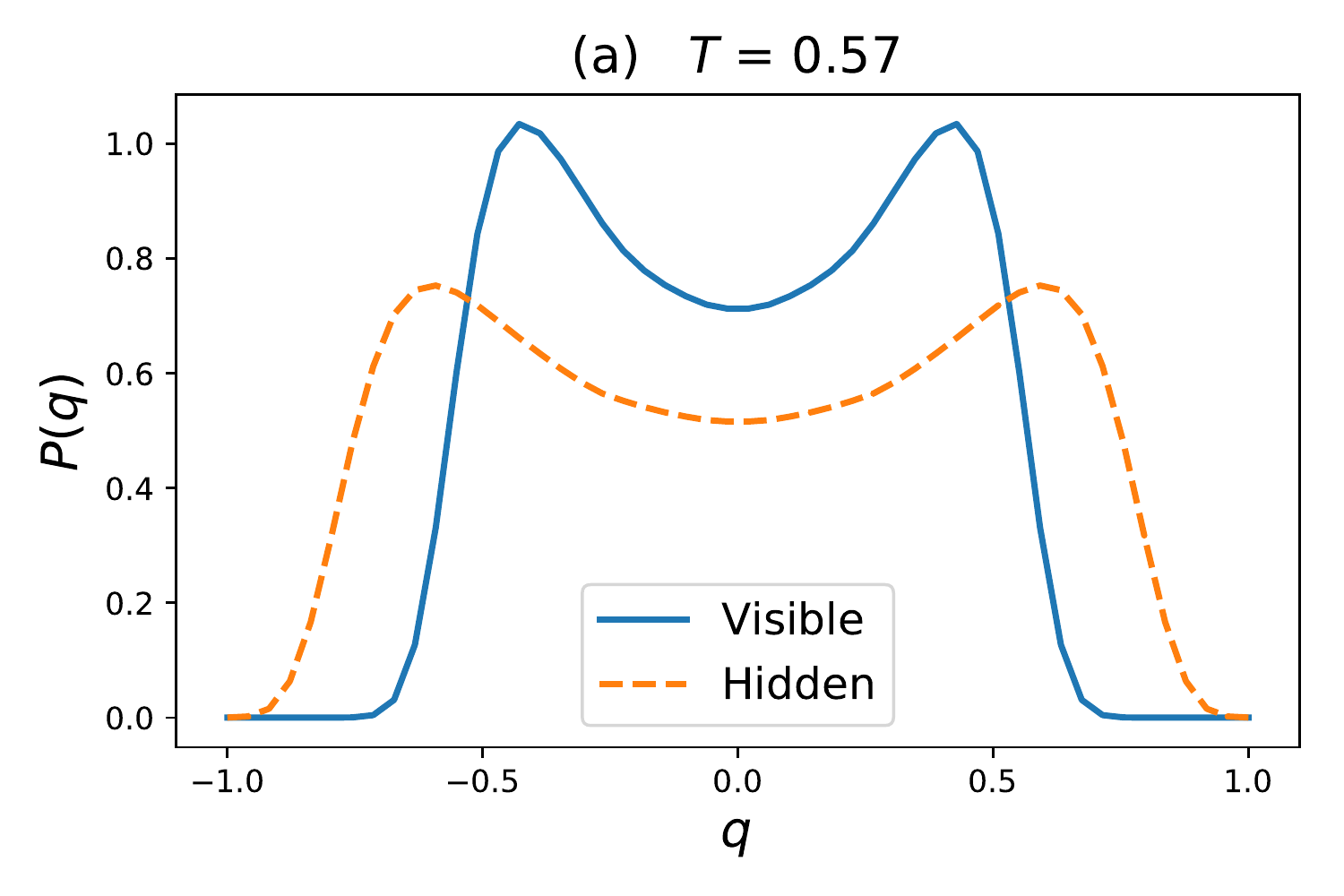}
\end{subfigure}%
\begin{subfigure}{.5\textwidth}
  \centering
\includegraphics[width=1.0\textwidth]{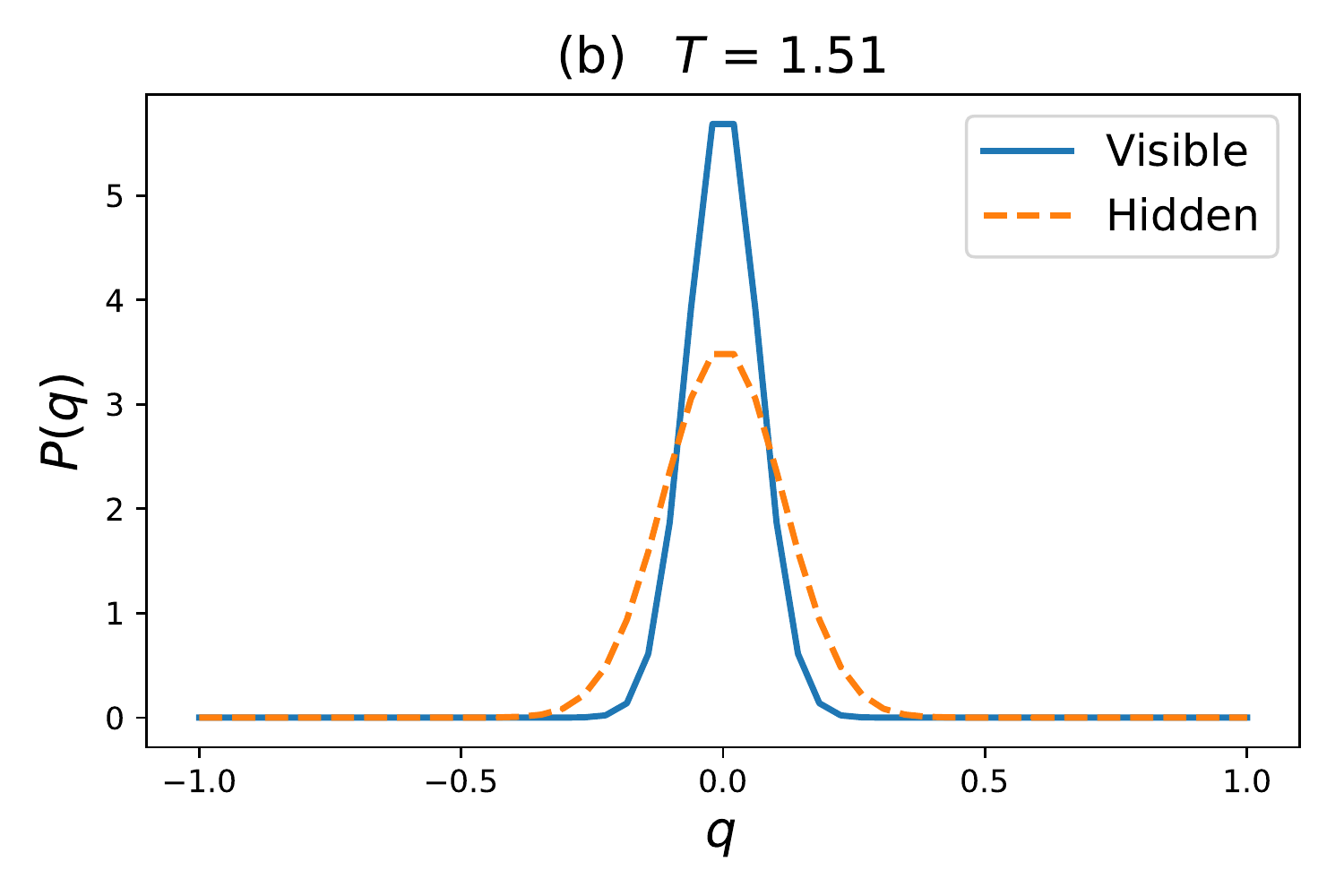}  
\end{subfigure}
\caption{\label{fig:overlap_distribution}Overlap distribution, $P(q)$, for the visible spins (blue, solid) and the hidden spins (orange, dashed), for (a): a temperature in the spin glass phase, and (b): a temperature in the paramagnetic phase. The distributions are estimated from $N=400$, $\alpha_h = 0.25$ MC samples using Parzen windows with a standard deviation of $0.05$.  In an infinite RSB analysis, one would expect delta function peaks at the maximal values of $q$, though these divergences are  difficult to capture with finite-$N$ simulations.}
\end{figure}

In Fig.~\ref{fig:overlap_distribution} we plot the overlap distributions $P(q)$ over the full range of $q \in [-1, 1]$ (see footnote~\ref{ftnt:Z2footnote}) for temperatures in both phases. For $T<T_c = 1$, the distribution is bimodal, with two well-separated peaks, indicating that replica symmetry has been broken. As the temperature is raised, the peaks move closer together, until they merge into a single, well-localized peak around $q=0$.

\begin{figure}
\centering
\begin{subfigure}{.3\textwidth}
  \centering
\includegraphics[width=1.1\textwidth]{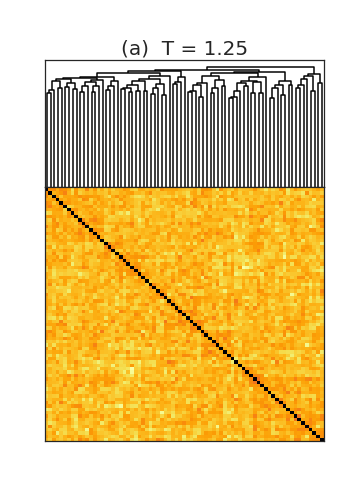}    
\end{subfigure}%
\begin{subfigure}{.3\textwidth}
  \centering
\includegraphics[width=1.1\textwidth]{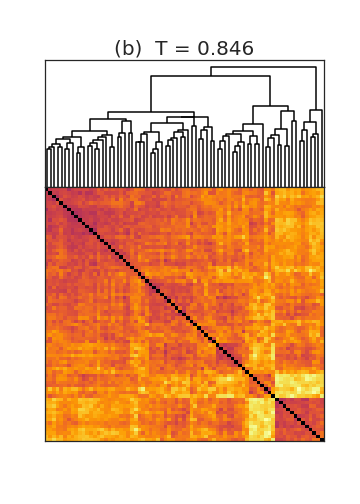}    
\end{subfigure}
\begin{subfigure}{.3\textwidth}
  \centering
\includegraphics[width=1.1\textwidth]{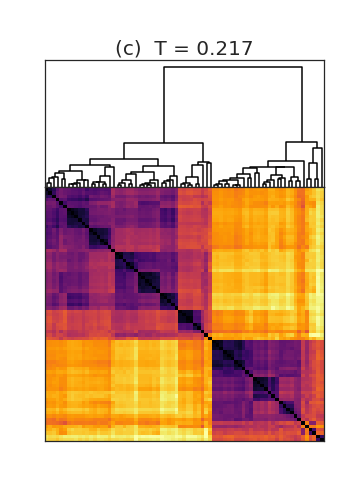}    
\end{subfigure}
\caption{\label{fig:ultrametric} Visual depictions of the ultrametric structure, or lack thereof, in the state space of the bipartite SK model for different temperatures and $w = 1$, $w_0 = 0$. The first temperature (a) corresponds to the paramagnetic (disordered) phase, and the others (b), (c), correspond to the spin glass (ordered) phase. Black pixels correspond to matrix entries with $d_{a b} = 0$, and the color then changes from dark purple to yellow as the distances increase. These plots were made for $N_v = N_h = 500$.}
  \vspace*{\floatsep}
\centering
\begin{subfigure}{.5\textwidth}
  \centering
\includegraphics[width=1.0\textwidth]{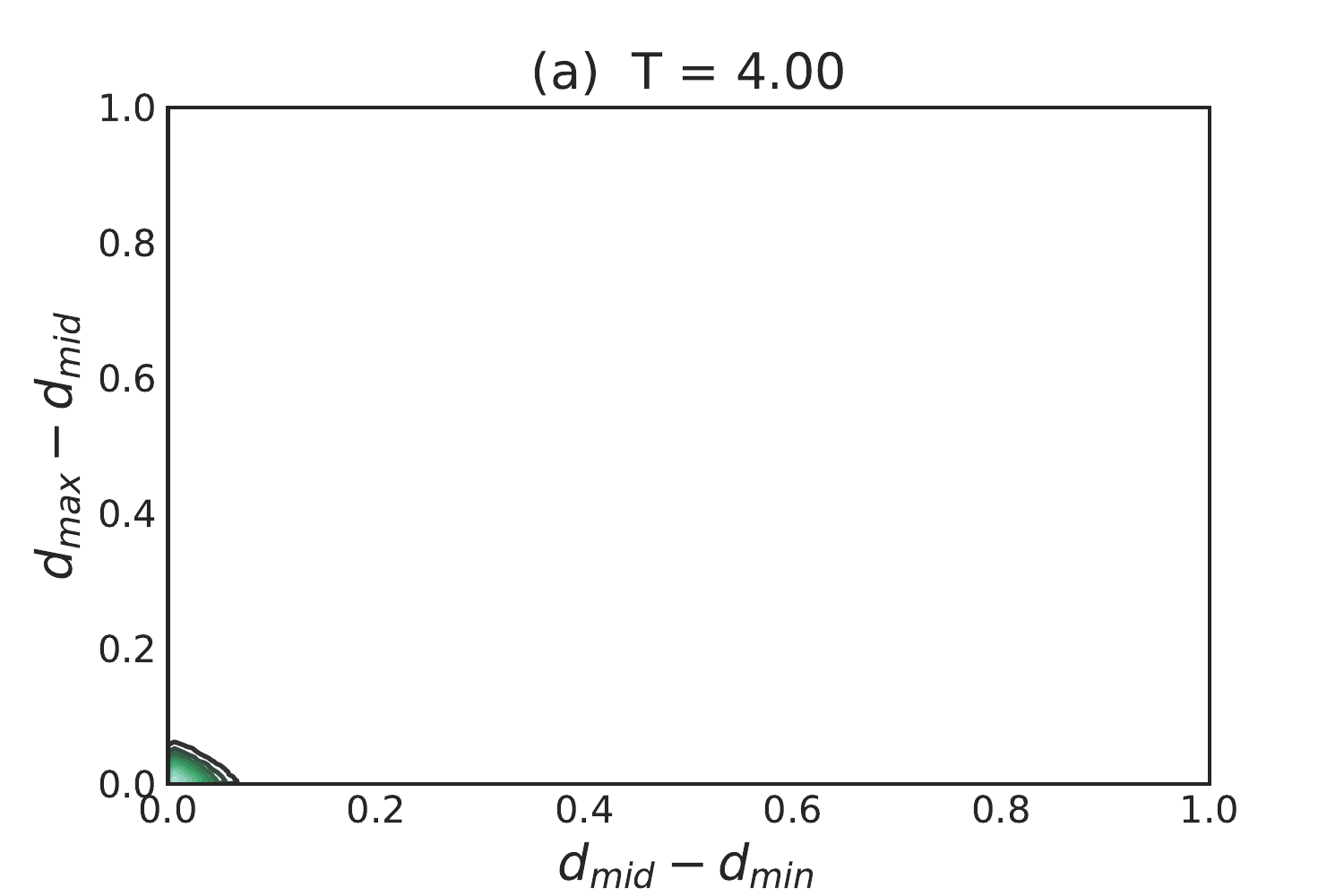}
\end{subfigure}%
\begin{subfigure}{.5\textwidth}
  \centering
\includegraphics[width=1.0\textwidth]{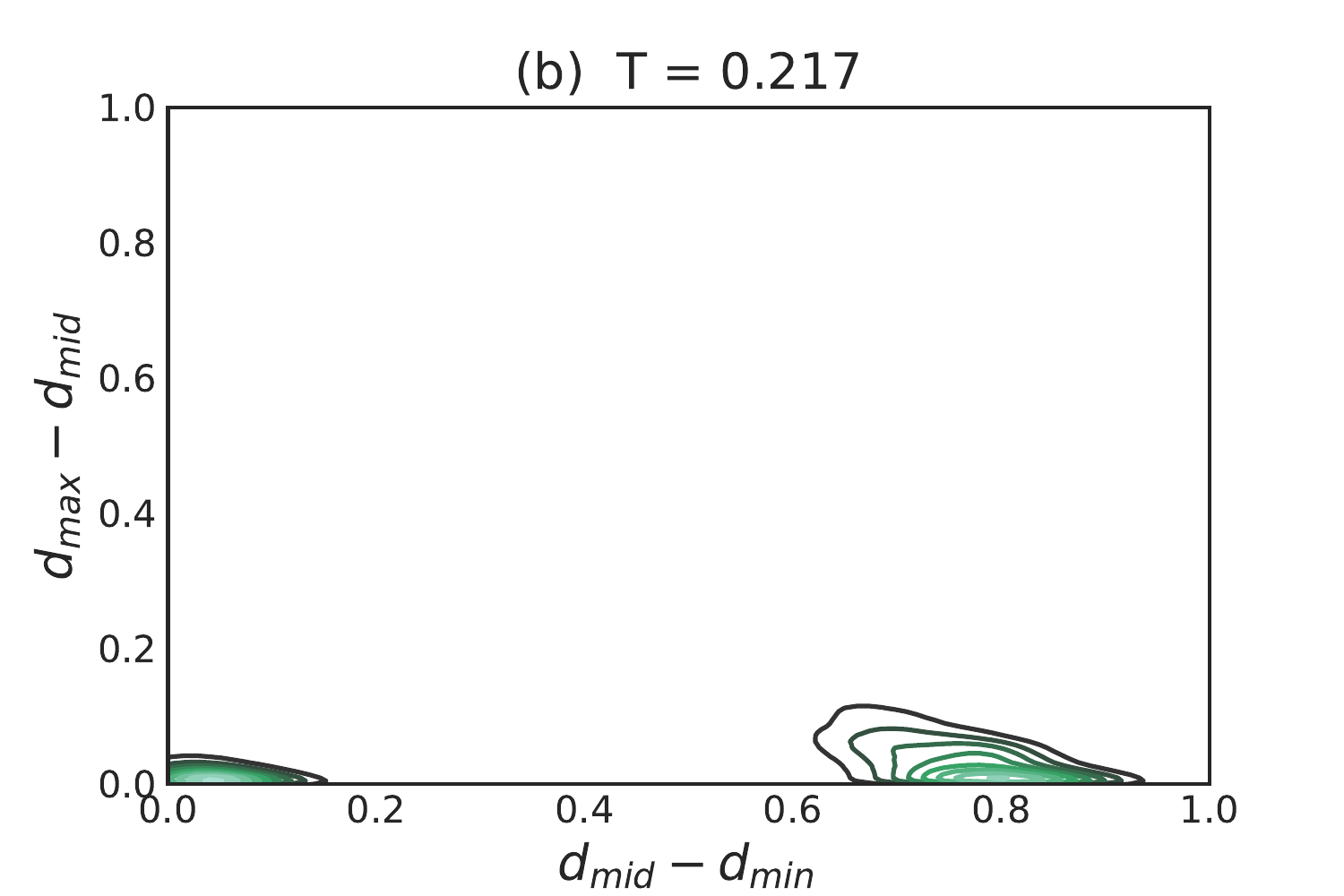}    
\end{subfigure}
\caption{\label{fig:distance_plot}
The probability density of triangle side lengths, for visible spins only, in the $(d_{\text{mid}} - d_{\text{min}}, d_{\text{max}} - d_{\text{mid}})$ plane for (a): the paramagnetic phase at $T=4.0$, and (b): the spin glass phase at $T=0.217$. Points which lie on the horizontal axis correspond to isosceles triangles, and the origin corresponds to equilateral triangles. In the spin glass phase, approximately 25\% of the points lie in the equilateral cluster near the origin, in agreement with the theoretical prediction Eq.~\ref{eq:3overlap_dist}. These plots were generated for $N_v = 400$, $N_h = 200$.}
\end{figure}

Lastly, we numerically investigated the ultrametric property of the spin glass phase. We created an interesting visualization of the relationship between the pure states very similar to Fig.~\ref{fig:parisi_ansatz}. Following \cite{hed2004lack, katzgraber2009ultrametricity}, we numerically simulated the model for only a single realization of the couplings $W_{ij}$, and obtained $M=150$ independent samples, for $N_v = N_h = 500$. For every pair of states $a$, $b$, the distance $d_{a b}$ can be computed, resulting in an $M\times M$ matrix. Using this distance matrix, we then applied a clustering algorithm to the states. Initially, each state belongs to its own cluster. Then, the two clusters which are closest together are joined into a new cluster. This process continues, with the definition of inter-cluster distances given by the average of all the distances between elements of each cluster. The process terminates when there is just a single cluster. This clustering process results in a tree, which is depicted in Fig.~\ref{fig:ultrametric} for various temperatures. The ordering of the endpoints, or leaves, of the tree provides a natural ordering of the states. Below the trees plotted in Fig.~\ref{fig:ultrametric} we also show the distance matrix $d_{a b}$, with the indices ordered as in the tree.  The resulting plots clearly illustrate a hierarchical relationship  between the states that in the spin glass phase, and absent in the paramagnetic phase. 

More insight into the degree of ultrametricity can be gleaned by considering the distribution of triangles formed by draws of three random pure states. For any triplet of states $a, b, c$, let the distances $0 \le d_{\text{min}} \le d_{\text{mid}} \le d_{\text{max}} \le 1$ be the sorted pairwise distances between these states. Various simple diagnostics of ultrametricity have been proposed in the literature, see for example \cite{marinari1998numerical, hed2004lack, katzgraber2009ultrametricity}. Unfortunately, it is difficult to construct diagnostics which are capable of distinguishing between so-called trivial ultrametricity, in which all the triangles are equilateral (which is the case for purely random spins) and non-trivial ultrametricity, in which a fraction of triangles are non-equilateral isosceles. We therefore directly examine the probability density of triangles, plotted in the $(d_{\text{mid}} - d_{\text{min}}, d_{\text{max}} - d_{\text{mid}})$ plane, as shown in Fig.~\ref{fig:distance_plot}.\footnote{After the first version of this paper appeared, we learned that this same approach was used in \cite{cogoni2017ultrametricity}.} If the states are organized ultrametrically, then they should all lie near the horizontal axes, which is indeed the case. In the spin glass phase there are clearly two clusters, one near the origin which corresponds to equilateral triangles, and a second cluster which corresponds to non-equilateral isosceles triangles. The equilateral cluster contains approximately $25\%$ of all points, in accord with the analytic prediction Eq.~\ref{eq:3overlap_dist}. In the paramagnetic phase there is only a single equilateral cluster.

\section{Application: Restricted Boltzmann Machines \label{sec:RBM}}
Restricted Boltzmann Machines (RBMs) are neural networks used in unsupervised machine learning applications, in which the goal is to  approximate the probability distribution that gave rise to a given data set. RBMs define a probability distribution over a space of two sets of variables $\bm{v}, \bm{h}$, called the visible and hidden neurons. The distribution is of the Boltzmann form, $p(\bm{v}, \bm{h}) = e^{-\beta H}/Z$, and for the case of Bernoulli RBMs, the Hamiltonian is of precisely the same expression as in the bipartite SK model, Eq.~\ref{eq:bipartiteSK}, with the crucial difference that the network parameters $W_{ij}, b_i^{(v)}, b_j^{(h)}$ are now determined by a learning procedure.\footnote{Although the Hamiltonians for both the bipartite SK model and the RBM are of the same form, the standard convention is that the RBM variables take on the values $v_i, h_j \in \{0,1\}$. We will continue with the spin convention used in Sec.~\ref{sec:bipartiteSK}, namely that $v_i, h_j \in \{-1,1\}$. The parameters for either convention are related through a simple linear transformation.}  

Given that the probability distribution defined by the RBMs is of the same parametric form as the equilibrium distribution of the bipartite SK model, it is interesting to consider to what extent RBMs exhibit spin-glass phenomena. The connection between neural networks and spin glasses has a long and rich history, including the seminal work of Amit et al \cite{amit1985spin, amit1987storing, amit1987statistical} on the problem of associative memory in Hopfield networks. This line of research has been recently extended to RBMs in \cite{barra2018phase, barra2017phase2}, where it was observed that RBMs are equivalent to generalized Hopfield networks and spin glass techniques were used to study the problem of memory retrieval. These works only studied the replica-symmetric case, leaving the replica symmetry broken analysis for later work. Other recent work on the thermodynamic properties of RBMs includes \cite{decelle2017spectral, decelle2018thermodynamics}, who focused on the singular value decomposition of the weight matrix, as well as \cite{huang2017statistical}, who used a message-passing-based approach. These works all assume replica symmetry; it would be very interesting to extend their results to the replica symmetry breaking case.

\begin{figure}
\centering
\begin{subfigure}{.5\textwidth}
  \centering
\includegraphics[width=1.0\textwidth]{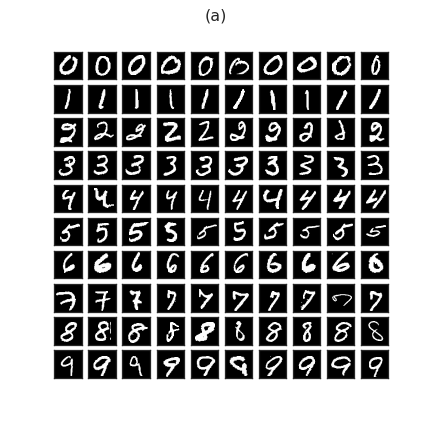}    
\end{subfigure}%
\begin{subfigure}{.5\textwidth}
  \centering
\includegraphics[width=1.0\textwidth]{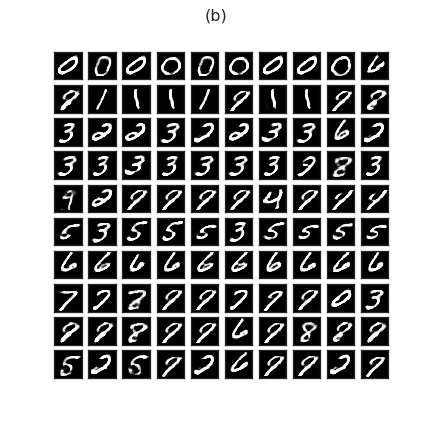}    
\end{subfigure}
\caption{\label{fig:MNIST} (a): A random selection of 100 MNIST images, with 10 examples of each digit. (b): Images generated from the RBM through Gibbs sampling, using the MNIST data shown in the left panel as initial seeds.}
\end{figure}

The network parameters for a practical RBM implementation will certainly not follow an iid Gaussian distribution as in Eq.~\ref{eq:gaussian_parameters}, so the results of Sec.~\ref{sec:bipartiteSK} should not be expected to directly apply to this more general case.\footnote{One interesting point on this topic is that the physics of the bipartite SK model is, to some weak extent, independent of the Gaussianity assumption. A universality result was proven for the SK model by \cite{talagrand2002gaussian, carmona2006universality} and recently extended to the bipartite SK model by \cite{genovese2012universality}. These authors showed that the behavior of the model is independent of the particular distribution of the parameters, provided the parameters are still iid and that the first two moments of the distributions are finite.} Additionally, it is very difficult to see how the general case can be treated analytically, without making some set of strongly simplifying assumptions that are not likely to hold in realistic scenarios.\footnote{Recently a new Boltzmann Machine has been proposed in \cite{krefl2017riemann} for which the partition function may be solved for in closed form. Although the usefulness of this new model for machine learning applications has yet to be evaluated, it is remarkable that a closed-form expression can be found for the partition function (and hence also the probabilities). It may be possible to derive analytic results for this model without assuming that the weights are iid.} We will therefore empirically investigate to what extent the trained RBM exhibits general spin glass behavior, including replica symmetry breaking and ultrametricity, and we will choose to focus on a RBM that has been trained on the MNIST character recognition data set. All results presented here are for a RBM with $N_v = 28^2 = 784$ visible and $N_h = 500$ hidden neurons, although we verified that the results are qualitatively the same for a range of $N_h$ values.\footnote{We have also repeated some of the analyses of this section for the Caltech 101 Silhouette data set \cite{Caltech101Silhouette}, and verified that the results are qualitatively unchanged.} In Fig.~\ref{fig:MNIST} we plot a selection of the original MNIST images (left) as well as a states generated through Gibb sampling (right), using the MNIST images as seeds.

\begin{figure}
\centering
\begin{subfigure}{.5\textwidth}
  \centering
\includegraphics[width=0.8\textwidth]{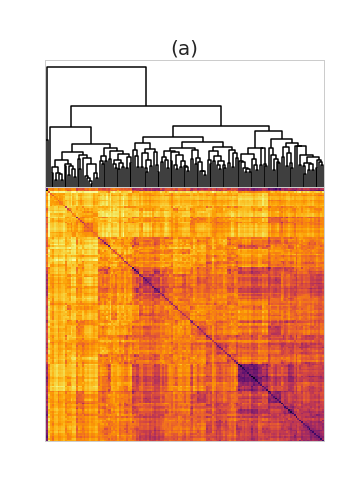}
\end{subfigure}%
\begin{subfigure}{.5\textwidth}
  \centering
\includegraphics[width=0.8\textwidth]{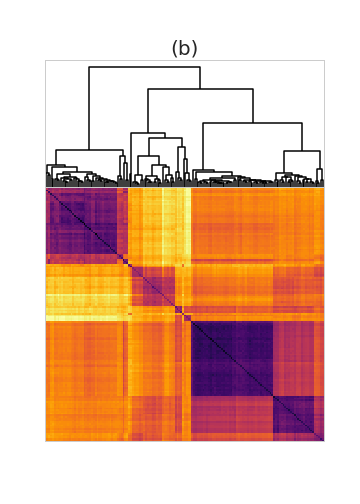}    
\end{subfigure}
\caption{\label{fig:ultrametric_MNIST} (a): The dissimilarity matrix and associated dendrogram for 150 images of the MNIST data, after converting to binary by rounding all non-zero pixel intensities to 1. (b): The same plot, made for the Gibbs samples drawn from the trained RBM, using the MNIST images as initial seeds. The distances are calculated according to the formula used in Fig.~\ref{fig:ultrametric}, namely ${d_{ab} = (1-N_v^{-1} \sum_i v_i^a v_i^b)/2}$.}
  \vspace*{\floatsep}
\centering
\begin{subfigure}{.5\textwidth}
  \centering
\includegraphics[width=1.0\textwidth]{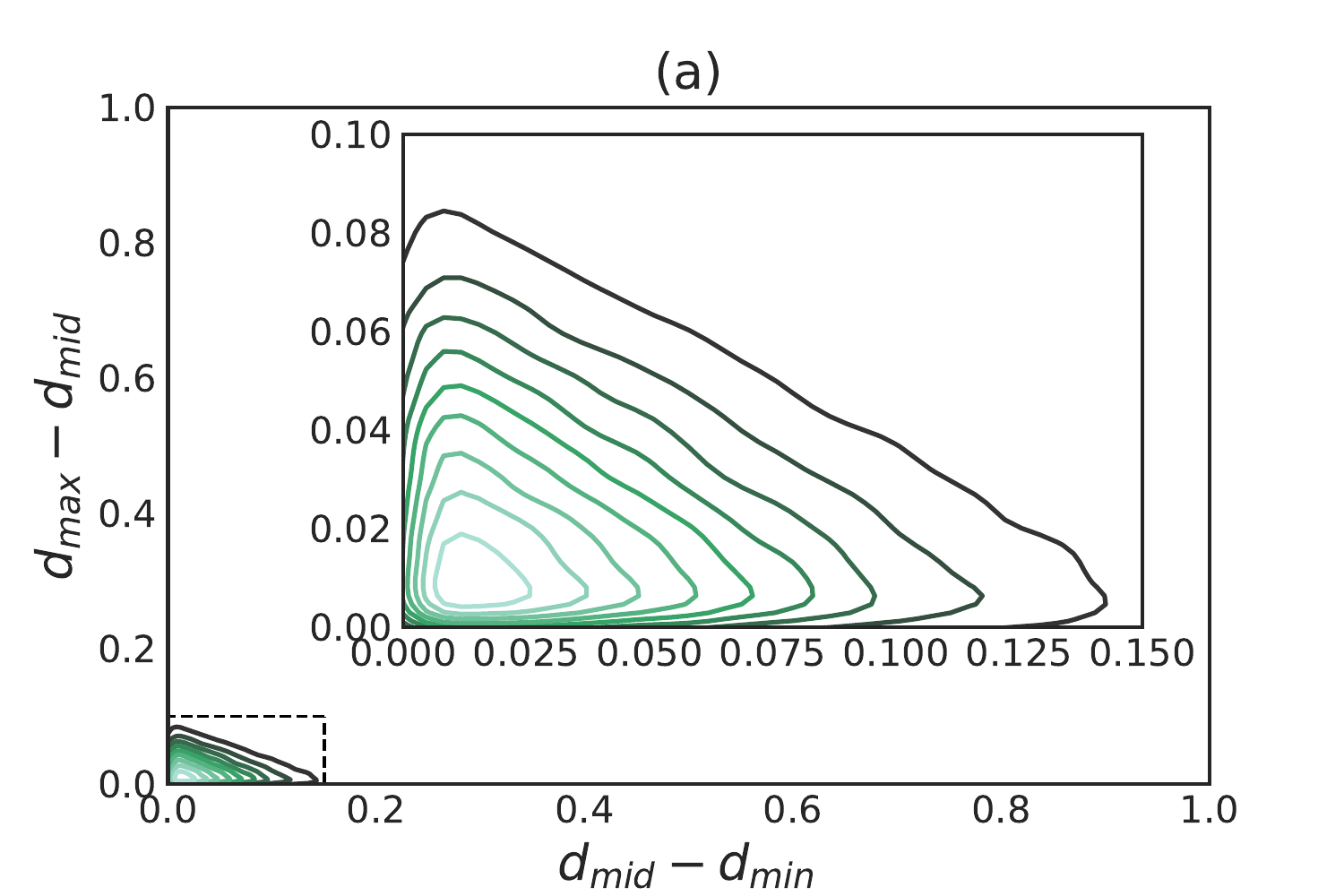}
\end{subfigure}%
\begin{subfigure}{.5\textwidth}
  \centering
\includegraphics[width=1.0\textwidth]{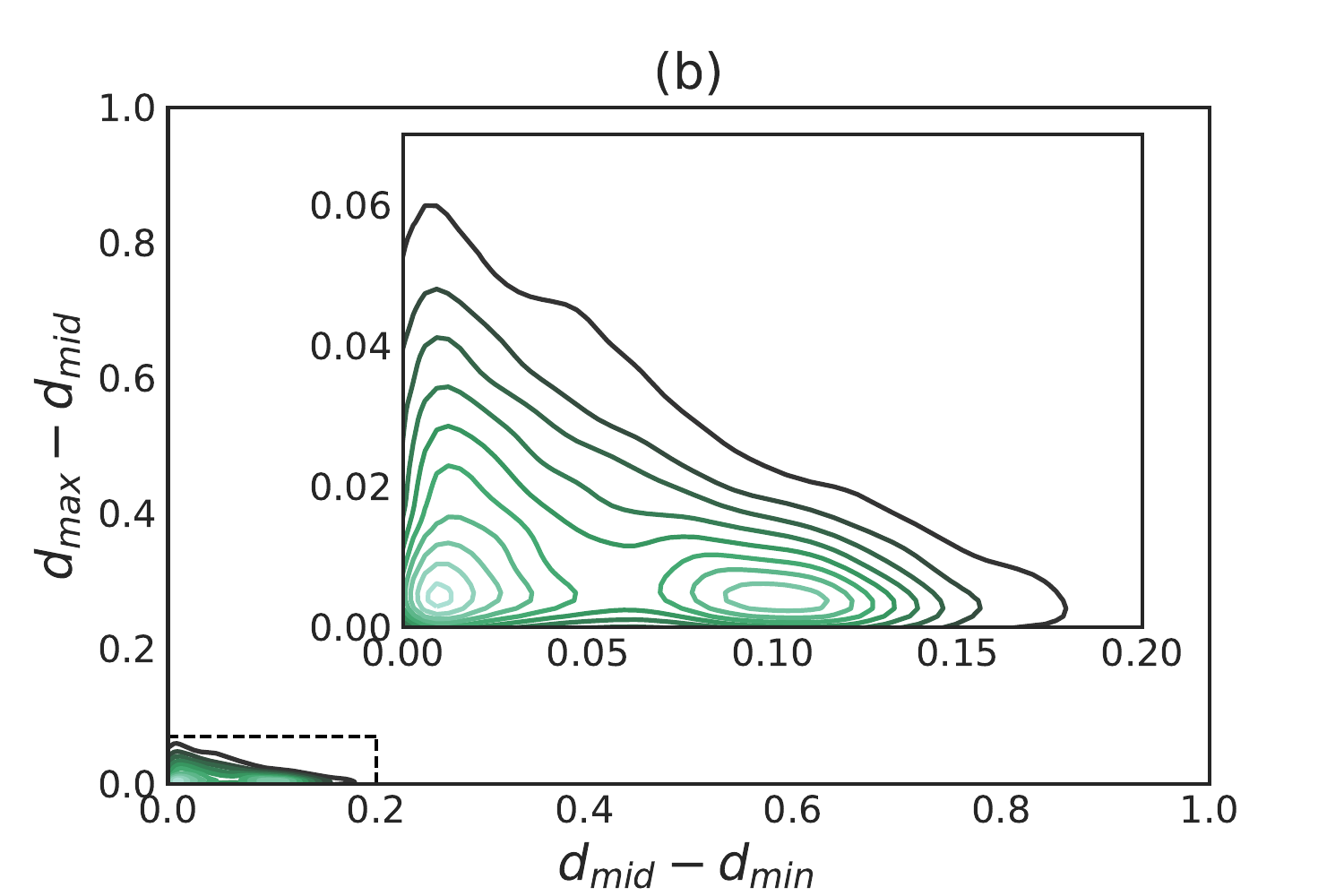} 
\end{subfigure}
\caption{\label{fig:MNIST_distance_plot} Distribution of triangle side lengths for the visible spins. (a): MNIST data, converted to binary by rounding all non-zero pixel intensities to 1. (b): Samples generated through Gibbs sampling a fully trained RBM. The insets show zoomed-in depictions of the distribution.}
\end{figure}

To gain some qualitative insight into the form of replica symmetry breaking that occurs (if any), in Fig.~\ref{fig:ultrametric_MNIST} we plot the dissimilarity matrix of Gibbs samples drawn from the trained RBM, again organizing the matrix entries according to a clustering analysis. We do the same for an arbitrary selection of MNIST images. Interestingly, the RBM samples lead to a plot that looks remarkably similar to the corresponding plot for the low-temperature spin glass phase of the bipartite SK model (the left-most image of Fig.~\ref{fig:ultrametric}). Both exhibit a strong blocking structure, and the RBM image looks like it could well have been generated by a spin glass simulation. In contrast, the MNIST plot exhibits some blocking structure, but it is much weaker, and is more similar to the near-critical spin glass image (the middle image of Fig.~\ref{fig:ultrametric}). The stark difference between the two plots is rather remarkable given the close visual similarity between the actual MNIST images and the RBM Gibbs samples, as shown in Fig.~\ref{fig:MNIST}.

To further investigate the extent to which the Gibbs samples exhibit ultrametricity, we next plot the distribution of triangles side lengths in Fig.~\ref{fig:MNIST_distance_plot}, in the same fashion as we did for the bipartite SK model in Fig.~\ref{fig:distance_plot}. We again show results for both the MNIST data and the Gibbs samples drawn from the RBM. The MNIST distribution seems to entirely consist of near-equilateral triangles, the same sort of trivial ultrametricity observed in the paramagnetic phase of the spin glass. The RBM distribution is different, however. Although there is not a clear separation between a cluster of equilateral triangles and a cluster of isosceles triangles, as was the case for the bipartite SK model, the contour plot reveals two distinct peaks. One peak is near the origin and corresponds to approximate equilateral triangles, and the other is centered around $(0,0.1)$ and corresponds to approximate isosceles triangles. If the peaks are separated by a vertical line at $d_{\text{mid}} - d_{\text{min}} \sim 0.06$, then the fraction of triangles which are approximately acute isosceles is roughly 40\%, compared to the 75\% for the bipartite SK and bipartite SK models, as implied by the Parisi ansatz.\footnote{The 40\% estimate is rough, and varies by a few percentage points depending on where the dividing line is drawn. The key point is that the fraction is far from the result for the bipartite SK model.} 

The above results suggest that RBMs trained on realistic data do in fact exhibit a significant degree of ultrametricity, just like spin glasses. This may seem somewhat natural, given that the RBM Hamiltonian is of the same form as in the bipartite SK model, and the results of Sec.~\ref{sec:bipartiteSK} which showed that the physics of the bipartite SK model are qualitatively the same as of the standard SK model. On the other hand, the universal approximation theorem for RBMs \cite{le2008representational} suggests that in the infinite-$N$ limit, a trained RBM should be able to precisely reproduce any training data distribution without changing (or introducing) ultrametric structure. We believe that there are two regimes of large-$N$ at play: for fairly large $N$, the spin-glass description seems to hold fairly well, but at even larger $N$ the universal approximation theorem eventually dominates and the trained RBM's ultrametric structure converges to that of the training distribution. An unusually high degree of ultrametric (or more generally, hierarchical) structure could conceivably even serve as a diagnostic for distinguishing natural data from the visually indistinguishable output of a well-trained RBM.

\section{Conclusion \label{sec:conclusion}}
We have considered a simple extension of the well-known Sherrington-Kirkpatrick (SK) model to bipartite graphs, and performed both a replica-symmetric and replica symmetry breaking analysis. In the latter case, we used the same Parisi ansatz originally introduced for the unipartite SK model for both the visible and hidden overlap matrices. Our combined analytical and numerical analyses revealed that the bipartite model exhibits the same general phenomena as the unipartite model, including the existence of a spin glass phase characterized by a continuous order parameter and an ultrametrically distributed space of pure states. As part of our analytical treatment we solved the RSB-1 equations, saving the case of infinite RSB for future work. Even with this somewhat crude approximation, we still found excellent agreement both with the results of MC simulations, and with the subleading correction to the large-$N$ optimal cost for a combinatorial graph partitioning problem.

We then applied these results to study the closely related family of neural networks known as Restricted Boltzmann Machines (RBMs). The Gibbs sampling outputs of the trained RBM exhibited properties which were qualitatively similar to the outputs of  Monte Carlo simulations of the bipartite SK model. Perhaps the most interesting aspect of the Parisi ansatz is the prediction of ultrametricity, which implies a hierarchical organization of states. Neural networks may be generally described as representing knowledge in a hierarchical fashion \cite{lecun2015deep}, and given the close connection between RBMs and the bipartite SK model, it is natural to wonder if the learned distribution exhibits ultrametricity. Interestingly, our empirical analysis revealed that the trained RBM exhibits a much stronger signature of ultrametricity than the training data.

Finally, we discuss some avenues for future work. First, it would be interesting to extend our results to Deep Boltzmann Machines (DBMs) with many hidden layers \cite{ruslan2009deep}. The Hamiltonian for a DBM with $L$ hidden layers is
\begin{equation}
H = - \sum_{I=1}^{L} \sum_{i_I = 1}^{N_I} \sum_{i_{I+1}=1}^{N_{I+1}} W_{i_I i_{I+1}}^{(I)} s_{i_I}^{(I)} s_{i_{I+1}}^{(I+1)} - \sum_{I=1}^{L+1} \sum_{i_I = 1}^{N_I} b_{i_I}^{(I)} s_{i_I}^{(I)} \, ,
\end{equation}
where the $s^{(I)}_{i_I} \in \{-1,1\}$ are spin variables. For $L=1$ this reduces to the expression for RBMs. If the weights $W^{(I)}$ and biases $b^{(I)}$ are again normally distributed, then the integral identity Eq.~\ref{eq:integralxform} may be used to derive an expression for the disorder-averaged free energy analogous to Eq.~\ref{eq:free_energy_no_simplification}, at which point either a replica symmetric or symmetry breaking analysis may be performed. It would be interesting to consider the $L \rightarrow \infty$ limit of such a network, and to compare the results of such a calculation with the behavior of DBMs with many layers trained on realistic data.

Lastly, it would also be very desirable to use spin glass theory to better understand the distribution of critical points of the loss function of RBMs, similar to what was done in \cite{choromanska2015loss} for feed-forward networks. Given the intractability of the RBM partition function for general values of the network parameters, this seems to be a difficult undertaking, even if strong simplifying assumptions are made regarding the training data. 

\subsubsection*{Acknowledgments}
This work is dedicated to the memory of Professor Joseph Polchinski. GSH would like to acknowledge many invaluable discussions, as well as a collaboration at an early stage of this project with Marco Caldarelli. GSH and EG would also like to thank Andrew Lohn and Osonde Osoba for useful discussions and for careful readings of this manuscript. The authors also thank Yan Fyodorov for bringing relevant literature to their attention and an anonymous referee for very helpful feedback. Part of this work was carried out while GSH was affiliated with the STAG Research Centre and the School of Mathematical Sciences at the University of Southampton, during which time he was supported by the STFC Ernest Rutherford grant ST/M004147/1.


\bibliography{paper}
\bibliographystyle{JHEP}

\end{document}